\documentclass[twocolumn,showpacs,aps,prd,nobibnotes,nofootinbib,floatfix,superscriptaddress,longbibliography]{revtex4-1}

\usepackage{amsmath,amssymb,bm,xfrac}
\usepackage{graphicx}
\usepackage[usenames]{xcolor}
\usepackage[normalem]{ulem}
\usepackage{xspace}
\usepackage{dsfont}
\usepackage{soul} 
\usepackage{dsfont}
\usepackage{multirow}
\usepackage{float}
\usepackage{cancel}
\usepackage[caption=false]{subfig}
\usepackage[colorlinks=true,allcolors=blue]{hyperref}

\newif\ifclean
\cleantrue
\newcommand{\rem}[1]{\ifclean\relax\else{\color{red}\sout{#1}}\fi}
\newcommand{\add}[1]{\ifclean#1\else{\color{blue}#1}\fi}
\newcommand{\rep}[2]{\rem{#1}\add{#2}}

\newcommand{\orcid}[1]{\href{https://orcid.org/#1}{\includegraphics[height=\fontcharht\font`\B]{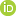}}}

\newcommand{\etal}{\emph{et~al}.\xspace}

\newcommand{\igm}{\texttt{IllinoisGRMHD}\xspace}
\newcommand{\ogm}{\texttt{OrigGRMHD}\xspace}
\newcommand{\baikal}{\texttt{Baikal}\xspace}

\newcommand{\harmnuc}{\texttt{HARM3D+NUC}\xspace}
\newcommand{\jupyter}{\texttt{Jupyter}\xspace}
\newcommand{\etk}{\texttt{Einstein Toolkit}\xspace}

\newcommand{\nrpy}{\texttt{NRPy+}\xspace}
\newcommand{\nrpyeos}{\texttt{NRPyEOS}\xspace}
\newcommand{\nrpyeoset}{\texttt{NRPyEOSET}\xspace}
\newcommand{\nrpyleakage}{\texttt{NRPyLeakage}\xspace}
\newcommand{\nrpyleakageet}{\texttt{NRPyLeakageET}\xspace}
\newcommand{\zelmanileak}{\texttt{ZelmaniLeak}\xspace}
\newcommand{\zelmanimone}{\texttt{ZelmaniM1}\xspace}
\newcommand{\eosomni}{\texttt{EOS\char`_Omni}\xspace}
\newcommand{\handoff}{\texttt{HandOff}\xspace}
\newcommand{\groned}{\texttt{GR1D}\xspace}
\newcommand{\spritz}{\texttt{Spritz}\xspace}
\newcommand{\grhydro}{\texttt{GRHydro}\xspace}

\newcommand{\carpet}{\texttt{Carpet}\xspace}

\newcommand{\compose}{\texttt{CompOSE}\xspace}
\newcommand{\thcleak}{\texttt{THC\char`_Leakage}\xspace}
\newcommand{\whiskythc}{\texttt{WhiskyTHC}\xspace}
\newcommand{\nubhlight}{\texttt{$\nu$bhlight}\xspace}
\newcommand{\spectre}{\texttt{SpeCTRE}\xspace}
\newcommand{\monegrey}{\texttt{M1Grey}\xspace}
\newcommand{\mpl}{\texttt{Matplotlib}\xspace}
\newcommand{\lorene}{\texttt{LORENE}\xspace}
\newcommand{\ahfd}{\texttt{AHFinderDirect}\xspace}
\newcommand{\qlm}{\texttt{QuasiLocalMeasures}\xspace}

\newcommand{\convec}{\bm{C}}
\newcommand{\primvec}{\bm{P}}
\newcommand{\fluxvec}{\bm{F}}
\newcommand{\sourcevec}{\bm{S}}
\newcommand{\tautilde}{\tilde{\tau}}
\newcommand{\rhostar}{\tilde{D}} 
\newcommand{\yestar}{\tilde{Y}_{\rm e}} 
\newcommand{\Msun}{M_{\odot}}
\newcommand{\Fdual}{{}^{*}\!F}
\newcommand{\rhob}{\rho_{\rm b}}
\newcommand{\ye}{\ensuremath{Y_{\rm e}}\xspace}
\newcommand{\mb}{m_{\rm b}}
\newcommand{\nb}{n_{\rm b}}
\renewcommand{\ne}{n_{\rm e}}

\newcommand{\nue}{\nu_{\rm e}}
\newcommand{\anue}{\bar{\nu}_{\rm e}}
\newcommand{\nux}{\nu_{x}}

\newcommand{\nui}{\nu_{i}}
\newcommand{\anui}{\bar{\nu}_{i}}
\renewcommand{\AA}{\mathcal{A}}
\newcommand{\RR}{\mathcal{R}}
\newcommand{\QQ}{\mathcal{Q}}
\newcommand{\sqrtgamma}{\sqrt{\gamma}}
\newcommand{\ee}{e^{-}}
\renewcommand{\ae}{e^{+}}

\newcommand{\rate}[3]{#1_{#3}^{#2}}

\newcommand{\mue}{\mu_{\rm e}}
\newcommand{\mun}{\mu_{\rm n}}
\newcommand{\mup}{\mu_{\rm p}}

\newcommand{\dx}{\Delta x}

\newcommand{\beq}{\begin{equation}}
\newcommand{\eeq}{\end{equation}}

\renewcommand{\eqref}[1]{Eq.~(\ref{#1})\xspace}

\newcommand{\eqrefalt}[1]{Eq.~\ref{#1}\xspace}

\newcommand{\figref}[1]{Fig.~\ref{#1}\xspace}

\newcommand{\figrefalt}[2]{Fig.~\ref{#1}{(#2)}\xspace}
\newcommand{\secref}[1]{Sec.~\ref{#1}\xspace}

\begin{document}

    \title{Addition of tabulated equation of state and neutrino leakage support to \igm}

\author{Leonardo~R.~Werneck~\orcid{0000-0002-4541-8553}}
\email{leonardo@uidaho.edu}
\affiliation{Department of Physics, University of Idaho, Moscow, ID 83843, USA}

\author{Zachariah~B.~Etienne~\orcid{0000-0002-6838-9185}}
\affiliation{Department of Physics, University of Idaho, Moscow, ID 83843, USA}
\affiliation{Department of Physics and Astronomy, West Virginia University, Morgantown, WV 26506}
\affiliation{Center for Gravitational Waves and Cosmology, West Virginia University, Chestnut Ridge Research Building, Morgantown, WV 26505}

\author{\\Ariadna~Murguia-Berthier~\orcid{0000-0003-2333-6116}}
\affiliation{Center for Interdisciplinary Exploration and Research in Astrophysics (CIERA), 1800 Sherman Ave., Evanston, IL 60201, USA}
\affiliation{NASA Einstein fellow}

\author{Roland~Haas~\orcid{0000-0003-1424-6178}}
\affiliation{National Center for Supercomputing Applications \& Department of Physics, University of Illinois, Urbana, IL 61801, USA}

\author{Federico~Cipolletta~\orcid{0000-0001-7894-1028}}
\affiliation{Center for Computational Relativity and Gravitation, Rochester Institute of Technology, Rochester, New York 14623, USA}

\author{Scott~C.~Noble~\orcid{0000-0003-3547-8306}}
\affiliation{Gravitational Astrophysics Lab, NASA Goddard Space Flight Center, Greenbelt, MD 20771, USA}

\author{Lorenzo~Ennoggi~\orcid{0000-0002-2771-5765}}
\affiliation{Center for Computational Relativity and Gravitation, Rochester Institute of Technology, Rochester, New York 14623, USA}

\author{Federico~G.~Lopez~Armengol~\orcid{0000-0002-4882-5672}}
\affiliation{Center for Computational Relativity and Gravitation, Rochester Institute of Technology, Rochester, New York 14623, USA}

\author{Bruno~Giacomazzo~\orcid{0000-0002-6947-4023}}
\affiliation{Università degli Studi di Milano - Bicocca, Dipartimento di Fisica G. Occhialini, Piazza della Scienza 3, I-20126 Milano, Italy}
\affiliation{INFN, Sezione di Milano-Bicocca, Piazza della Scienza 3, I-20126 Milano, Italy}
\affiliation{INAF, Osservatorio Astronomico di Brera, Via E. Bianchi 46, I-23807 Merate (LC), Italy}

\author{Thiago~Assumpção~\orcid{0000-0002-3419-892X}}
\affiliation{Department of Physics and Astronomy, West Virginia University, Morgantown, WV 26506}
\affiliation{Center for Gravitational Waves and Cosmology, West Virginia University, Chestnut Ridge Research Building, Morgantown, WV 26505}

\author{Joshua~Faber~\orcid{0000-0003-1724-3474}}
\affiliation{Center for Computational Relativity and Gravitation, Rochester Institute of Technology, Rochester, New York 14623, USA}
\affiliation{School of Mathematical Sciences, Rochester Institute of Technology, Rochester, New York 14623, USA}
\affiliation{School of Physics and Astronomy, Rochester Institute of Technology, Rochester, New York 14623, USA}

\author{Tanmayee~Gupte}
\affiliation{Center for Computational Relativity and Gravitation, Rochester Institute of Technology, Rochester, New York 14623, USA}
\affiliation{School of Physics and Astronomy, Rochester Institute of Technology, Rochester, New York 14623, USA}

\author{Bernard~J.~Kelly~\orcid{0000-0003-2666-6268}}
\affiliation{Center for Space Sciences and Technology, University of Maryland Baltimore County, 1000 Hilltop Circle Baltimore, MD 21250, USA}
\affiliation{Gravitational Astrophysics Lab, NASA Goddard Space Flight Center, Greenbelt, MD 20771, USA}
\affiliation{Center for Research and Exploration in Space Science and Technology, NASA Goddard Space Flight Center, Greenbelt, MD 20771, USA}

\author{Julian~H.~Krolik~\orcid{0000-0002-2995-7717}}
\affiliation{Physics and Astronomy Department, Johns Hopkins University, Baltimore, MD 21218, USA}

\begin{abstract}
  We have added support for realistic, microphysical, finite-temperature equations of state (EOS) and neutrino physics via a leakage scheme to \igm, an open-source GRMHD code for dynamical spacetimes in the \etk. These new features are provided by two new, \nrpy-based codes: \nrpyeos, which performs highly efficient EOS table lookups and interpolations, and \nrpyleakage, which implements a new, AMR-capable neutrino leakage scheme in the \etk. We have performed a series of strenuous validation tests that demonstrate the robustness of these new codes, particularly on the Cartesian AMR grids provided by \carpet. Furthermore, we show results from fully dynamical GRMHD simulations of single unmagnetized neutron stars, and magnetized binary neutron star mergers. This new version of \igm, as well as \nrpyeos and \nrpyleakage, is pedagogically documented in \jupyter notebooks and fully open source. The codes will be proposed for inclusion in an upcoming version of the \etk.
\end{abstract}

\maketitle

\section{Introduction}
\label{sec:introduction}

Magnetized fluid flows in dynamical spacetimes are a key driver of multimessenger phenomena, a prominent example of which was the gravitational-wave signal GW170817~\cite{TheLIGOScientific:2017qsa} and the coincident short gamma-ray burst GRB170817A~\cite{GBM:2017lvd}, originating from a binary system of two merging neutron stars (NSs). Self-consistent simulations of these systems require software capable of modeling the diverse physics of the problem, from relativistic general relativistic magnetohydrodynamics (GRMHD) fluid flows, to the hot degenerate matter described by a microphysical, finite-temperature equation of state (EOS), to the changes in matter composition and energy due to the emission and absorption of neutrinos and photons, to the rapidly changing spacetime dynamics involved in the merger and black hole (BH) formation.

Given the high demand for accurate simulations of these systems, it is unsurprising that multiple groups have developed their own codes, some of which specialize in GRMHD for stationary spacetimes, which can be used e.g., for studying merger remnants~\cite{Gammie:2003rj,Noble:2005gf,Noble:2008tm,Murguia-Berthier:2021tnt,Anderson:2006ay}; while others are intended to model more generic GRMHD flows in dynamical spacetimes, which can be used for inspiral, merger, and post-merger dynamics~\cite{Bruegmann:2006ulg,OConnor:2009iuz,Thierfelder:2011yi,Giacomazzo:2007ti,Cerda-Duran:2008qfl,Kiuchi:2012qv,Dionysopoulou:2012zv,Radice:2012cu,Radice:2013hxh,Radice:2013xpa,Moesta:2013dna,White:2015omx,Etienne:2015cea,2020ascl.soft04003E,Kidder:2016hev,10.1145/3330345.3330346,2018JCoPh.375.1365F,Most:2019kfe,Cipolletta:2019geh,Cipolletta:2020kgq,2020ApJS..249....4S,Mewes:2020vic,tichy2020numerical}. Having many codes means that a variety of algorithmic choices are made during their development, some of which impact performance, some the code's suitability to model certain physical systems, and others the physical realism of the simulations.

Examples of differences that affect a code's performance and/or its ability to accurately model certain physical systems include the numerical resolution and adopted coordinate system (e.g., Cartesian, spherical,  etc.); how spatial derivatives are represented numerically (e.g., finite difference, finite volume, discontinuous Galerkin, spectral, or even hybrid methods); the choice of EOS; and how neutrino effects are modeled; just to name a few.

Regarding the EOS, a common choice made by many codes is a simple ideal gas EOS, accounting for temperature effects through a thermal contribution~\cite{1993A&A...268..360J}. Others improve the description of the cold (as opposed to thermal) portion of the EOS by adopting the so-called piecewise polytropic model (see e.g.,~\cite{Read:2008iy} and references therein). Finally, some codes adopt microphysical, finite-temperature EOS tables constructed using data from astrophysical observations and nuclear physics experiments, like the ones available in the \compose~\cite{Typel:2013rza,Oertel:2016bki,Typel:2022lcx} and Stellar Collapse~\cite{stellarcollapse_website} databases. These tables provide the best description of nuclear matter available to date.

Regarding neutrino effects, one may consider the general relativistic radiation magnetohydrodynamics (GRRMHD) equations and model neutrino transport via Monte Carlo methods---by far the most computationally expensive option. These methods try to solve the seven-dimensional Boltzmann's equation by grouping particles into \emph{packets} that approximate the distribution function of neutrinos at random points. We mention here in particular the works of Foucart \etal~\cite{Foucart:2020qjb,Foucart:2021mcb,Foucart:2021ikp} and Miller \etal~\cite{Miller:2019gig,Miller:2019dpt}, which have used this technique in the context of binary neutron star (BNS) mergers. It is important to note that this method becomes prohibitively expensive in the optically thick regime (high densities and temperatures), requiring e.g., enforcing a hard ceiling on the value of the absorption opacity of the fluid~\cite{Foucart:2021mcb}.
 
 Another approximate method for modeling neutrino physics is moment-based radiation transport. In this technique, the Boltzmann equation is recast as a $3+1$ system~\cite{1981MNRAS.194..439T,Shibata:2011kx}, which is then solved using similar numerical techniques to those for the GRMHD equations. Unlike the GRMHD equations, however, the system cannot be closed with an EOS, making its accuracy dependent on the choice of closure (see e.g.,~\cite{Richers:2020ntq}). This method has also been used in many studies, including core-collapse~\cite{Obergaulinger:2014nna,OConnor:2014sgn,Kuroda:2015bta,OConnor:2015rwy,Roberts:2016lzn,Skinner:2018iti,Glas:2018oyz,Rahman:2019yxy,Laiu:2021pha} and BNS~\cite{Foucart:2015vpa,Foucart:2015gaa,Foucart:2016rxm,Radice:2021jtw,Sun:2022vri}.

Leakage schemes are perhaps the most popular approach for modeling neutrino physics~\cite{vanRiper:1981mko,Ruffert:1995fs,Rosswog:2002rt,Rosswog:2003rv,Sekiguchi:2010ep,Sekiguchi:2011zd}. In this approach, experimental data are used to parameterize analytic formulas for the neutrino emission and cooling rates in terms of the optical depths, resulting in a computationally inexpensive algorithm that has been adopted rather broadly~\cite{OConnor:2009iuz,Ott:2012kr,Neilsen:2014hha,Radice:2016dwd,Siegel:2017jug,Endrizzi:2019trv,Murguia-Berthier:2021tnt}. \add{It is important to state that leakage scheme typically neglect the absorption of neutrinos and therefore do not account for heating and lepton number changes due to neutrinos in the ejecta~\cite{Just:2021cls}.}

While these studies differ in how neutrinos are modeled, one aspect most share in common is that thHey were performed using software that is not freely available to everyone. Exceptions include the GRMHD codes \groned~\cite{OConnor:2009iuz,OConnor:2014sgn}, \whiskythc~\cite{Radice:2012cu,Radice:2013hxh,Radice:2013xpa}, \grhydro~\cite{Moesta:2013dna}, \igm~\cite{Etienne:2015cea,2020ascl.soft04003E}, \spectre~\cite{Kidder:2016hev,spectrecode}, and \spritz~\cite{Cipolletta:2019geh,Cipolletta:2020kgq,giacomazzo_bruno_2020_4350072}; the neutrino leakage codes \zelmanileak~\cite{OConnor:2009iuz,Ott:2012kr} and \thcleak~\cite{Radice:2016dwd}; the moment-based radiation transport codes \zelmanimone~\cite{Roberts:2016lzn} and \monegrey~\cite{Kidder:2016hev,spectrecode}; and the GRRMHD Monte Carlo code \nubhlight~\cite{Miller:2019gig}.

Here we introduce a major update to \igm---a concise open-source rewrite of the Illinois numerical relativity group's GRMHD code~\cite{Duez:2005sf} (henceforth \ogm), which exists within the \etk. This new version, which is also open-source~\cite{igm_github}, supports both finite-temperature, microphysical EOSs---via a new \nrpy~\cite{Ruchlin:2017com} module called \nrpyeos~\cite{igm_github}---and neutrino physics via a leakage scheme---using the recently developed \nrpy-based code \nrpyleakage~\cite{igm_github}. This updated version will be proposed for inclusion in a future \etk release.

For well over a decade, both \ogm and \igm have been used to model a plethora of astrophysical scenarios, including magnetized binary neutron stars (BNS)~\cite{Liu:2008xy,Paschalidis:2012ff,Ruiz:2017inq,Tsokaros:2019anx,Raithel:2021hye,Sun:2022vri,Armengol:2021mbt}, binary BH-NS~\cite{Etienne:2007jg,Etienne:2008re,Etienne:2011ea,Etienne:2012te,Paschalidis:2014qra,Ruiz:2018wah,Ruiz:2020elr}, BH accretion disks~\cite{Farris:2011vx,Farris:2012ux,Gold:2013zma,Gold:2014dta,Khan:2018ejm,EventHorizonTelescope:2019pcy,Wessel:2020hvu}, binary white dwarf-NS~\cite{Paschalidis:2010dh,Paschalidis:2011ez}, rotating NSs~\cite{Etienne:2006am,Espino:2019xcl}, gravitational collapse of supermassive stars~\cite{Sun:2017voo}, magnetized Bondi accretion~\cite{Etienne:2010ui}, and magnetized hypermassive neutron stars (HMNS)~\cite{Duez:2005cj,Duez:2006qe,Liu:2007cf,Shibata:2005mz,Shibata:2006hr,Stephens:2006cn,Ruiz:2020via}, to name a few. We also highlight \texttt{Frankfurt}/\igm~\cite{Most:2019kfe}, whose feature set exceeds that of the original \igm, but like \ogm is currently a closed-source code.

In contrast to these codes, \igm is open source and part of the \etk~\cite{Loffler:2011ay}, and in this work we introduce \etk \emph{thorns} (or modules) for \nrpyeos and \nrpyleakage, named \nrpyeoset and \nrpyleakageet, respectively.%
\footnote{We will refer to these as simply \nrpyeos and \nrpyleakage unless discussing thorn-exclusive features.}
This new version of \igm, along with \nrpyeos and \nrpyleakage, will be proposed for inclusion in a future release of the \etk, but in the meantime all codes are freely available for download at~\cite{igm_github}.

\rem{\nrpyeos is a pedagogically documented and infrastructure-agnostic \nrpy module that generates table interpolation routines based on the \etk's core EOS driver thorn \eosomni, which is itself based on the original code by O'Connor \& Ott~\cite{eosdrivercxx_repo}.
\nrpyeos provides a clean and clear user interface, generating specialized routines that compute only needed hydrodynamic quantities, avoiding unnecessary interpolations and thus greatly increasing the overall performance of GRMHD simulations that make use of tabulated EOSs.
When computing the neutrino opacities and emission and cooling rates, for example, \nrpyleakage requires five table quantities: \mbox{$\left(\mue,\mun,\mup,X_{\rm n},X_{\rm p}\right)$}, which are the chemical potentials of the electron, neutron, and proton, and the neutron and proton mass fractions, respectively. The general-purpose routine \texttt{EOS\char`_Omni\char`_full} is the only one available in \eosomni to compute such quantities. This routine, however, interpolates a total of 17 table quantities, 11 of which are unused for our purposes, resulting in it being twice as slow as the specialized routine generated by \nrpyeos.}

\rem{\nrpyleakage, like \nrpyeos, is also pedagogically documented and infrastructure-agnostic. It implements the neutrino leakage scheme of~\mbox{\cite{Ruffert:1995fs,Galeazzi:2013mia,Siegel:2017jug}}, while also considering neutrino production from nucleon-nucleon Bremsstrahlung as in~\mbox{\cite{Burrows:2004vq,OConnor:2011pxx}}. In contrast to \zelmanileak, which computes the optical depths using a spherical ray-by-ray integration algorithm, optical depths are computed using the more generic local algorithm of~\mbox{\cite{Neilsen:2014hha,Siegel:2017jug,Murguia-Berthier:2021tnt}}. Further, its \etk version has been carefully designed to work seamlessly with the Cartesian AMR grids provided by \carpet~\cite{Schnetter:2003rb}. The user is given the option of disabling any of the neutrino production channels in \nrpyleakage, thus allowing the code to be fully compatible with the neutrino leakage scheme of \harmnuc~\cite{Murguia-Berthier:2021tnt}, which does not include nucleon-nucleon Bremsstrahlung.}

\rem{Compatibility with \harmnuc remains a high priority for the authors so that in the post-merger phase, after the spacetime has become sufficiently stationary, the \handoff code~\cite{Armengol:2021mbt} is used to transfer simulation data from \igm to \harmnuc. We thus replace the Cartesian grid used by \igm---which is not ideal for modeling accretion disks, as the plasma flows obliquely to coordinate lines and has its angular momentum spuriously sapped by numerical errors---by the spherical-like coordinate system used by \harmnuc. Because this coordinate system has been optimized to model BH accretion disks, it enables us to accurately and reliably evolve the remnant spacetime for the relatively long time scales associated with multimessenger astronomy.}

This paper is organized as follows. \secref{sec:basic_equations} provides an overview of the mathematical formulation of the GRMHD equations and of the neutrino leakage scheme. \secref{sec:numerical_methods} describes the numerical methods and technical aspects of our codes. \secref{sec:results} contains a series of challenging validation tests of the code, as well as results from simulations of single NSs, and magnetized, equal-mass BNS systems, with eventual BH formation. \secref{sec:conclusions} contains closing remarks and plans for future work. 

\section{Basic equations}
\label{sec:basic_equations}

Throughout the paper Greek letters $\mu,\nu,\rho,\ldots$ are used to denote spacetime indices (range 0--3) and lowercase Roman letters $i,j,k,\ldots$ to denote spatial indices (range 1--3), assuming Einstein summation convention. Unless stated otherwise, geometrized units \mbox{$G = c = 1$} are adopted, additionally assuming that \mbox{$\Msun=1$}. Temperatures are measured in ${\rm MeV}$.

The spacetime evolution is governed by Einstein's equations,
\begin{equation}
  G^{\mu\nu} = 8\pi T^{\mu\nu}\;,\label{eq:Einstein}
\end{equation}
where $G^{\mu\nu}$ is the Einstein tensor and $T^{\mu\nu}$ is the total stress-energy tensor. As written, these equations are not in a form immediately suitable for numerical integration. One such form is the initial value formulation built upon first splitting $g_{\mu\nu}$ into the $3+1$ Arnowitt--Deser--Misner (ADM) form~\cite{Arnowitt:1962hi}
\begin{equation}
  ds^{2} = -\alpha^{2}dt^{2}+\gamma_{ij}\bigl(dx^{i}+\beta^{i}dt\bigr)\bigl(dx^{j}+\beta^{j}dt\bigr)\;,\label{eq:three_plus_one_metric}
\end{equation}
where $\alpha$, $\beta^{i}$, and $\gamma_{ij}$ are the lapse function, the shift vector, and the metric, all defined on spatial hypersurfaces of constant coordinate time $t$. With this decomposition Einstein's equations can be split into sets of hyperbolic (time-evolution) and elliptic (constraints) PDEs, similar to Maxwell's equations in differential form (see e.g.,~\cite{Baumgarte:2010ndz} for a pedagogical review). The resulting ADM evolution and constraint equations are in fact not numerically stable, and must be reformulated further. The Baumgarte--Shapiro--Shibata--Nakamura (BSSN) formulation~\cite{Nakamura:1987zz,Shibata:1995we,Baumgarte:1998te} is one such reformulation in which additional auxiliary and conformal variables are introduced in order to make the resulting set of equations strongly hyperbolic, enabling stable, long-term time integration of Einstein's equations on the computer.

The remainder of this section is dedicated to the GRMHD formulation used by \igm, as well as an overview of the neutrino leakage scheme adopted by \nrpyleakage.

\subsection{General relativistic magnetohydrodynamics}
\label{sec:grmhd}

\rep{Assuming infinite conductivity (ideal MHD, where \mbox{$u_{\mu}F^{\mu\nu}=0$}), t}{T}he GRMHD equations \add{with neutrino leakage} can be written as\rem{ the system:}
\begin{align}
  \nabla_{\mu}\left(\nb u^{\mu}\right) &= 0\;,\label{eq:baryon_number_conservation}\\
  \nabla_{\mu}\left(\ne u^{\mu}\right) &= \RR\;,\label{eq:lepton_number_conservation}\\
  \nabla_{\mu}T^{\mu\nu} &= \QQ u^{\nu}\;,\label{eq:enmom_conservation}\\
  \nabla_{\mu}\Fdual^{\mu\nu} &= 0\;,\label{eq:maxwell}
\end{align}
which are the conservation of baryon number, conservation of lepton number, conservation of energy-momentum, and homogeneous Maxwell's equations, respectively. In the above, $\nb$ ($\ne$) is the baryon (\rep{electron}{lepton}) number density, $u^{\mu}$ is the fluid four-velocity, $\mb$ is the baryon mass, \mbox{$\Fdual^{\mu\nu}=(1/2)\tilde{\epsilon}^{\mu\nu\rho\sigma}F_{\rho\sigma}$} is the dual of the Faraday tensor $F^{\mu\nu}$, and $\tilde{\epsilon}^{\mu\nu\rho\sigma}$ is the Levi-Civita tensor.
\add{Ideal MHD (\mbox{$u_{\mu}F^{\mu}=0$}) is assumed throughout.}
The source terms $\RR$ and $\QQ$ account for changes in lepton number and energy, respectively, due to the emission and absorption of neutrinos. The precise form of $\RR$ and $\QQ$ is detailed in the next section on neutrino leakage.

The energy-momentum tensor is assumed to be that of a perfect fluid, plus an electromagnetic contribution, given by
\begin{equation}
  T^{\mu\nu} = \left(\rhob h + b^{2}\right)u^{\mu}u^{\nu} + \left(P + \frac{b^{2}}{2}\right)g^{\mu\nu} - b^{\mu}b^{\nu}\;,
\end{equation}
where \mbox{$\rhob = \mb\nb$} is the baryon density, \mbox{$h = 1 + \epsilon + P/\rhob$} is the specific enthalpy, $\epsilon$ is the specific internal energy, $P$ is the fluid pressure, $b^{\mu}=(4\pi)^{-1/2}B^{\mu}_{(u)}$ is the rescaled 4-magnetic field in the fluid frame, where
\begin{align}
  B^{0}_{(u)} &= u_{i}B^{i}/\alpha\;,\\
  B^{i}_{(u)} &= \bigl(B^{i}/\alpha + B^{0}_{(u)}u^{i}\bigr)/u^{0}\;,
\end{align}
and $B^{i}$ is the magnetic field in the frame normal to the hypersurface.

To source Einstein's equations $T^{\mu\nu}$ must be updated from one time step to the next, which requires evolving the matter fields in time. To accomplish this, \mbox{Eqs.~(\ref{eq:baryon_number_conservation})--(\ref{eq:maxwell})} are rewritten in conservative form:
\begin{equation}
  \partial_{t}\convec + \partial_{i}\fluxvec^{i} = \sourcevec\;,\label{eq:grmhd_conservative_form}
\end{equation}
where $\fluxvec$ and $\sourcevec$ are the flux and source vectors, respectively. Furthermore, $\convec=\convec(\primvec)$ is the vector of \emph{conservative} variables, with $\primvec$ the vector of \emph{primitive} variables. \igm adopts the Valencia formalism \cite{Anton:2005gi,Banyuls:1997zz}, whereby
\begin{equation}
  \primvec = \left[
  \begin{array}{c}
  \rhob\\
  \ye\\
  T\\
  P\\
  v^{i}\\
  B^{i}
  \end{array}
  \right].\label{eq:prims}
\end{equation}
Here, $\ye\equiv\ne/\nb$ is the electron fraction, $T$ is the temperature, and \mbox{$v^{i}=u^{i}/u^{0}$} is the fluid three-velocity. Notice that this choice of primitive three-velocity differs from the Valencia three-velocity $v_{(n)}^{i}$ used in other codes (see e.g.,~\cite{Baiotti:2010zf,Baiotti:2010zf,Giacomazzo:2007ti,Moesta:2013dna,Cipolletta:2019geh,Cipolletta:2020kgq}); \igm adopts the 3-velocity that appears in the induction equation \eqref{eq:induction}. These two velocities are related via
\begin{equation}
  v_{(n)}^{i} = \alpha^{-1}\bigl(v^{i}+\beta^{i}\bigr)\;.
\end{equation}

The conservative variables can be written in terms of the primitive variables as
\begin{equation}
  \renewcommand{\arraystretch}{1.1}
  \convec \!=\!
  \left[
    \begin{array}{c}
      \rhostar\\
      \yestar\\
      \tautilde\\
      \tilde{S}_{i}\\
      \tilde{B}^{i}
    \end{array}
  \right]
  \!\equiv\!
  \sqrtgamma
  \left[
    \begin{array}{c}
      D\\
      D\ye\\
      \tau\\
      S_{i}\\
      B^{i}
    \end{array}
  \right]
  \!\equiv\!
  \sqrtgamma
  \left[
    \begin{array}{c}
      W\rhob\\
      D\ye\\
      \alpha^{2}T^{00} - D\\
      \alpha T^{0}_{\ i}\\
      B^{i}
    \end{array}
  \right],
  \label{eq:conservs}
  \renewcommand{\arraystretch}{1}
\end{equation}
where $\gamma = \det(\gamma_{ij})$ and $W=\alpha u^{0}$ is the Lorentz factor.

There are many numerical advantages associated with writing the GRMHD equations in conservative form. First, ignoring neutrino effects, the source terms $\sourcevec$ vanish in flat space with Cartesian coordinates and this form of the equations, when combined with an appropriate finite-volume scheme, guarantees conservation of total rest mass, lepton number, energy, and momentum to roundoff error. Second, conservation is guaranteed up to errors in $\sourcevec$ when the sources are nonzero. Third, \igm adopts finite-volume methods, which are superior at handling ultrarelativistic flows as compared to easier-to-implement and more-efficient artificial viscosity schemes~\cite{Marti:1991wi,Anninos:2002gz}. Fourth, this conservative formulation enables us to easily use a high-resolution shock-capturing (HRSC) scheme, designed to minimize Gibbs' oscillations near shocks, to name a few advantages.

An important consideration when implementing a GRMHD code is how to handle the magnetic induction equation, obtained from the spatial components of \eqref{eq:maxwell}. In conservative form, this equation may be written as
\begin{equation}
  \partial_{t}\tilde{B}^{i} + \partial_{j}\left(v^{j}\tilde{B}^{i} - v^{i}B^{j}\right) = 0\;.\label{eq:induction}
\end{equation}
If \eqref{eq:induction} were to be propagated forward in time, evaluating its spatial partial derivatives without special techniques, violations of the ``no magnetic monopoles'' condition (which follows from the time component of \eqrefalt{eq:maxwell})
\begin{equation}
  \partial_{i}\tilde{B}^{i} = 0\;,\label{eq:no_monopoles}
\end{equation}
will grow with each iteration. Ensuring this constraint remains satisfied, particularly on AMR grids, is a nontrivial endeavor. We adopt the same strategy as the one used in~\cite{Etienne:2010ui,Etienne:2011re}, which is to evolve the electromagnetic (EM) four-potential $\AA_{\mu}$ instead of the magnetic fields directly. In this case, numerical errors associated with evolving $\AA_{\mu}$ do not impact violations of \eqref{eq:no_monopoles}, as the magnetic field is computed as the ``curl'' (i.e., a Newtonian curl with definition appropriately generalized for GR) of the vector potential, and the divergence of the curl is by definition zero (here to roundoff error, once a numerical approximation for partial derivative is chosen).

The $3+1$ decomposition of the vector potential gives (see e.g.,~\cite{Baumgarte:2021skc} for a pedagogical review)
\begin{equation}
  \AA_{\mu} = n_{\mu}\Phi + A_{\mu}\quad \text{and}\quad \tilde{B}^{i} = \epsilon^{ijk}\partial_{j}A_{k}\;,\label{eq:em_four_potential}
\end{equation}
where $n_{\mu}$ is the unit vector normal to the spatial hypersurface, $A_{\mu}$ is the purely spatial (i.e., $n^{\mu}A_{\mu}=0$) magnetic potential, $\Phi$ is the electric potential, and $\epsilon^{ijk}$ is the totally antisymmetric Levi-Civita symbol, with $\epsilon^{123}=1$. Thus \eqref{eq:induction} becomes
\begin{equation}
  \partial_{t}A_{i} = \epsilon_{ijk}v^{j}\tilde{B}^{k} - \partial_{i}\left(\alpha\Phi - \beta^{j}A_{j}\right)\;.\label{eq:induction_magnetic_potential}
\end{equation}

We fix the EM gauge by adopting a covariant version of the ``generalized Lorenz gauge condition'',
\begin{equation}
  \nabla_{\mu}\AA^{\mu} = \xi n_{\mu}\AA^{\mu}\;,
\end{equation}
which was first introduced by the Illinois relativity group in~\cite{Etienne:2011re,Farris:2012ux}. Here $\xi$ is a parameter with units of inverse length, chosen so that the Courant--Friedrichs--Lewy (CFL) condition is always satisfied. Typically $\xi$ is set to $1.5/\Delta t_{\rm max}$, where $\Delta t_{\rm max}$ is the time step of the coarsest refinement level (see \cite{Mewes:2020vic} for further details). This gauge condition results in the additional evolution equation
\begin{equation}
  \partial_{t}\tilde{\Phi} + \partial_{j}\left(\alpha\sqrt{\gamma}A^{j} - \beta^{j}\tilde{\Phi}\right) = -\xi\alpha\tilde{\Phi}\;,\label{eq:lorenz_gauge_evolution}
\end{equation}
where $\tilde{\Phi} \equiv \sqrt{\gamma}\Phi$.

Except for Eqs.\,(\ref{eq:induction_magnetic_potential}) and (\ref{eq:lorenz_gauge_evolution}), the remaining GRMHD equations are evolved using \eqref{eq:grmhd_conservative_form} and a HRSC scheme, as described in~\cite{Etienne:2015cea}. For completeness, the remaining components of the flux vector are given by
\begin{equation}
  \bm{F}^{i}
  =
  \left[
    \begin{array}{c}
      \rhostar v^{i}\\
      \yestar v^{i}\\
      \alpha^{2}\sqrt{\gamma}T^{0i}-\rhostar v^{i}\\
      \alpha\sqrt{\gamma}T^{i}_{\ j}
    \end{array}
  \right]\;,
  \label{eq:grmhd_fluxes}
\end{equation}
and those of the source vector are given by
\begin{equation}
  \bm{S}
  =
  \left[
    \begin{array}{c}
      0\\
      \alpha\sqrt{\gamma}\RR\\
      s + \alpha\sqrt{\gamma}W\QQ\\
      \alpha\sqrt{\gamma}\left(\tfrac{1}{2}T^{\mu\nu}g_{\mu\nu,i} + \QQ u_{i}\right)
    \end{array}
  \right]\;,
  \label{eq:grmhd_sources}
\end{equation}
where
\begin{equation}
  \begin{split}
    s = \alpha\sqrt{\gamma}\Bigl[\bigl(T^{00}\beta^{i}\beta^{j}+2T^{0i}&\beta^{j}+T^{ij}\bigr)K_{ij}\\
    &-\left(T^{00}\beta^{i}+T^{0i}\right)\partial_{i}\alpha\Bigr]\;,
  \end{split}
\end{equation}
and $K_{ij}$ is the extrinsic curvature.

Specifying the matter EOS closes this system of equations. To this end, \igm supports both analytic, hybrid EOSs~\cite{1993A&A...268..360J} as well as microphysical, finite-temperature, fully tabulated EOSs.

For tabulated EOSs, hydrodynamic quantities are given as functions of the density $\rhob$, the electron fraction $\ye$, and the temperature $T$. As needed, the temperature can be recovered from the pressure, specific internal energy, or entropy by using either a Newton--Raphson method or bisection, with the latter yielding superior results in highly degenerate regions of parameter space.

\add{These table operations are handled by \nrpyeos, a pedagogically documented and infrastructure-agnostic \nrpy module that generates table interpolation routines based on the \etk's core EOS driver thorn \eosomni, which is itself based on the original code by O'Connor \& Ott~\cite{eosdrivercxx_repo}.
\nrpyeos provides a clean and clear user interface, generating specialized routines that compute only needed hydrodynamic quantities, avoiding unnecessary interpolations and thus greatly increasing the overall performance of GRMHD simulations that make use of tabulated EOSs.

A concrete example is the computation of the neutrino opacities and emission and cooling rates, for which \nrpyleakage requires five table quantities: $\left(\mue,\mun,\mup,X_{\rm n},X_{\rm p}\right)$, which are the chemical potentials of the electron, neutron, and proton, and the neutron and proton mass fractions, respectively. The general-purpose routine \texttt{EOS\char`_Omni\char`_full} is the only one available in \eosomni to compute such quantities. This routine, however, interpolates a total of 17 table quantities, 11 of which are unused for our purposes, resulting in it being twice as slow as the specialized routine generated by \nrpyeos.}

For hybrid EOSs, hydrodynamic quantities are analytic functions of the density and the specific internal energy. The pressure is given by
\begin{equation}
  P(\rhob,\epsilon) = P_{\rm cold}(\rhob) + P_{\rm thermal}(\rhob,\epsilon)\;,
\end{equation}
where
\begin{equation}
  P_{\rm thermal}(\rhob,\epsilon) = \left(\Gamma_{\rm th}-1\right)\rhob\left[\epsilon-\epsilon_{\rm cold}(\rhob)\right]\;,
\end{equation}
accounts for thermal effects, with $\Gamma_{\rm th}$ a constant parameter that determines the conversion efficiency of kinetic to thermal energy at shocks, and $P_{\rm cold}(\rhob)$ and $\epsilon_{\rm cold}(\rhob)$ are computed assuming either a gamma-law or piecewise polytropic EOS (see e.g.,~\cite{Read:2008iy}).

\subsection{Neutrino leakage}
\label{sec:neutrino_leakage}

\nrpyleakage---a new \add{infrastructure-agnostic} neutrino leakage code generated by \nrpy and fully documented in pedagogical \jupyter notebooks---enables us to incorporate basic neutrino physics in our simulations. Our implementation follows the prescription of~\cite{Ruffert:1995fs,Galeazzi:2013mia,Siegel:2017jug} to compute the neutrino number and energy emission rates, as well as the neutrino opacities. We also consider nucleon-nucleon Bremsstrahlung~\cite{Burrows:2004vq} following the \zelmanileak code~\cite{OConnor:2009iuz,Ott:2012kr,OConnor:2011pxx}. Unlike~\cite{Ruffert:1995fs} and \zelmanileak, however, we adopt a local, iterative algorithm in order to compute the neutrino optical depths following~\cite{Neilsen:2014hha,Siegel:2017jug} (see also~\cite{Murguia-Berthier:2021tnt})\xspace\rep{. This algorithm enables}{, yielding} far more efficient computations of optical depths than that implemented in \zelmanileak when the modeled system is far from spherical symmetry. \add{Further, the \etk version of the code has been carefully designed to work seamlessly with the Cartesian AMR grids provided by \carpet~\cite{Schnetter:2003rb}.}

Neutrinos are accounted for via the following reactions: \mbox{$\beta$-processes}, i.e., electrons ($\ee$) being captured by protons ($p$) and positrons ($\ae$) being captured by neutrons ($n$),
\begin{align}
  \ee + p &\to n + \nue\;,\\
  \ae + n &\to p + \anue\;;
\end{align}
electron-positron pair annihilation,
\begin{equation}
  \ee + \ae \to \nui + \anui\;;
\end{equation}
transverse plasmon ($\tilde\gamma$) decay,
\begin{equation}
  \tilde\gamma \to \nui + \anui\;;
\end{equation}
and nucleon-nucleon Bremsstrahlung,
\begin{equation}
  N + N \to N + N + \nui + \anui\;.
\end{equation}
In the reactions above $\nui=\left\{\nue,\nu_{\mu},\nu_{\tau}\right\}$ are the electron, muon, and tau neutrinos, with $\anui$ their antineutrinos, and $N$ heavy nuclei. Our implementation assumes that the contributions from heavy lepton neutrinos and antineutrinos are all the same, and we use the notation $\nux$ to refer to any one species throughout.

The production of neutrinos via these processes lead to changes in the electron fraction and energy of the system, which are accounted for in the source terms $\RR$ and $\QQ$ of Eqs.\,(\ref{eq:lepton_number_conservation}) and (\ref{eq:enmom_conservation}). More specifically,
\begin{align}
  \RR &= - \rate{\RR}{\nue}{\rm eff}
         + \rate{\RR}{\anue}{\rm eff}\;,
  \label{eq:R_source}\\
  \QQ &= - \rate{\QQ}{\nue}{\rm eff}
         - \rate{\QQ}{\anue}{\rm eff}
         - 4\rate{\QQ}{\nux}{\rm eff}\;,
  \label{eq:Q_source}
\end{align}
where the effective rates are given by
\begin{align}
  \rate{\RR}{\nui}{\rm eff} &= \rate{\RR}{\nui}{\rm free}\left[1 + t^{\nui,\RR}_{\rm diff}/t^{\nui,\RR}_{\rm free}\right]^{-1}\;,\label{eq:effective_emission_rates}\\
  \rate{\QQ}{\nui}{\rm eff} &= \rate{\QQ}{\nui}{\rm free}\left[1 + t^{\nui,\QQ}_{\rm diff}/t^{\nui,\QQ}_{\rm free}\right]^{-1}\;,\label{eq:effective_cooling_rates}
\end{align}
and the diffusion time scales using
\begin{equation}
  t^{\nui,j}_{{\rm diff}} = D_{\rm diff}\left(\kappa^{\nui}_{{\rm t},j}\right)^{-1}\left(\tau^{\nui}_{j}\right)^{2}\;,\label{eq:diffusion_time_scale}
\end{equation}
where $\tau^{\nui}_{j}$ are the neutrino optical depths, $\kappa^{\nui}_{{\rm t},j}$ the total neutrino transport opacity \add{(see \secref{sec:computation_of_optical_depths})}, \mbox{$D_{\rm diff}=6$}~\cite{Rosswog:2003rv,OConnor:2009iuz,Siegel:2017jug,Murguia-Berthier:2021tnt}, and $j=\RR,\QQ$. The free emission time scales are given by
\begin{equation}
  t^{\nui,\RR}_{\rm free} = \frac{n_{\nui}}{\rate{\RR}{\nui}{\rm free}}\;,\quad t^{\nui,\QQ}_{\rm free} = \frac{\varepsilon_{\nui}}{\rate{\QQ}{\nui}{\rm free}}\;,
\end{equation}
where $n_{\nui}$ and $\varepsilon_{\nui}$ are the neutrino number and energy density, respectively, and the total free emission and cooling rates are given by
\begin{align}
  \rate{\RR}{\nui}{\rm free}  &= \rate{\RR}{\nui}{\beta}
                               + \rate{\RR}{\nui,\anui}{\rm Pair}
                               + \rate{\RR}{\nui,\anui}{\rm Plasmon}
                               + \rate{\RR}{\nui,\anui}{\rm Bremss}\;,
  \label{eq:total_nui_emission_rate}\\
  \rate{\QQ}{\nui}{\rm free}  &= \rate{\QQ}{\nui}{\beta}
                               + \rate{\QQ}{\nui,\anui}{\rm Pair}
                               + \rate{\QQ}{\nui,\anui}{\rm Plasmon}
                               + \rate{\QQ}{\nui,\anui}{\rm Bremss}\;.
  \label{eq:total_nui_cooling_rate}
\end{align}
Note that $\beta$-processes only contribute when $\nui=\left\{\nue,\anue\right\}$. 

For small densities and temperatures---the optically thin regime---the optical depths vanish and the medium is essentially transparent to neutrinos. Diffusion occurs on much shorter time scales than free-streaming, and therefore the effective rates become the free ones. For large densities and temperatures---the optically thick regime---optical depths are large and the neutrinos interact with the matter, such that diffusion happens on long time scales and free-streaming on short ones due to the increase in the emission and cooling rates, implying \mbox{$\rate{\RR}{\nui}{\rm eff}\to n_{\nui}/t^{\nui,\RR}_{{\rm diff}}$} and \mbox{$\rate{\QQ}{\nui}{\rm eff}\to\varepsilon_{\nui}/t^{\nui,\QQ}_{{\rm diff}}$}. \add{The leakage scheme provides a smooth interpolation between the diffusive and free-streaming regimes (see~\cite{Ruffert:1995fs} for details).} We postpone the \rep{details of}{discussion on} how the optical depths are computed in \nrpyleakage until \secref{sec:computation_of_optical_depths}.

\section{Numerical methods}
\label{sec:numerical_methods}

Most of the core numerical algorithms in \igm remain the same as in the original version announced in 2015; these algorithms were reimplemented from \ogm as they were found most robust when modeling a large variety of astrophysical scenarios. Core algorithms, as described in~\cite{Etienne:2015cea}, include the HRSC scheme, the Harten--Lax--van Leer approximate Riemann solver~\cite{doi:10.1137/1025002}, the piecewise parabolic method~\cite{Colella:1982ee} used to reconstruct the primitive variables at the cell interfaces, the staggering of the electric and magnetic potentials and of the magnetic field, the algorithm to compute the magnetic fields from the magnetic potential, the Runge--Kutta time integration, and outflow boundary conditions.

Key algorithmic changes introduced here include an updated conservative-to-primitive infrastructure, which has been expanded to include new effective 1D routines that are well-suited for tabulated EOSs, and the interface with the newly developed codes \nrpyeos and \nrpyleakage. We reserve the remainder of this section to discussing these updates in detail.

\subsection{Conservative-to-primitive recovery}
\label{sec:conservative_to_primitive}

The energy-momentum tensor in the GR field equations and the GRMHD equations are written as functions of the primitive variables. So after updating the conservative variables at each time iteration, the primitive variables (``primitives") must be computed from the conservative variables (``conservatives"). This is a nontrivial step, as there are no algebraic expressions to compute the conservatives from the primitives, requiring the implementation of a root-finding algorithm to solve a set of coupled nonlinear equations.

As numerical errors---such as truncation error originating from spatial and temporal finite differencing, as well as interpolation and prolongation operations---can cause the conservative variables to stray away from their valid range, sometimes this inversion becomes impossible. Because of this, we perform a series of checks on conservative variables to ensure they are valid before attempting to recover the primitive variables. We refer the reader to Appendix A of~\cite{Etienne:2011ea} for details on how these bounds are checked and enforced.

For gamma-law and hybrid EOSs, the primary primitive recovery routine used in \igm is the 2D scheme of Noble~\etal~\cite{Noble:2005gf} (henceforth ``Noble 2D").%
\footnote{The dimensionality of the scheme is associated with how many equations are used to recover the primitive variables: 1D schemes use one equation and one unknown, 2D schemes use two equations and two unknowns, etc.}
We have also implemented the 1D scheme of the same reference, as well as 1D routines that replace the energy by the entropy, which is passively advected alongside the other variables assuming a conservation equation, as in~\cite{Noble:2008tm}. The user is then given the option to use one or more of these last three routines as backups to the Noble 2D one.

The entropy routines perform well in regions of high magnetization and low densities, where the other two can be less robust, but we stress that they should only be used as backups, as the entropy is not conserved at shocks and therefore cannot always be reliably used to recover the primitives. If the Noble 2D and backup routines are unable to recover the primitives, a final backup routine due to Font~\etal~\cite{Font:1998hf} is used, for which the pressure is reset to its cold value and thus an inversion is guaranteed (see Appendix A of~\cite{Etienne:2011ea} for more details).

For tabulated EOSs, Newton--Raphson-based routines become very sensitive to the initial guesses provided for the primitive values. \igm does not keep track of the values of the primitives at the previous time step, making it difficult to use routines such as the Noble 2D and the closely related routines by Ant\'on~\etal~\cite{Anton:2005gi}, Giacomazzo \& Rezzolla~\cite{Giacomazzo:2007ti}, and Cerd\'a-Dur\'an~\etal~\cite{Cerda-Duran:2008qfl} (see also~\cite{Murguia-Berthier:2021tnt}).

As reviewed in~\cite{Siegel:2017sav}, some routines require better initial guesses than others. In particular, we find that the 1D routines of Neilsen~\etal~\cite{Neilsen:2014hha} and Palenzuela~\etal~\cite{Palenzuela:2015dqa}, as well as the one by Newman \& Hamlin~\cite{NewmanHamlin}, which only require an initial guess for the temperature, are very robust at recovering primitive variables even for relatively poor initial guesses. Our implementation of these routines is based on the open-source infrastructure by Siegel~\cite{grmhd_con2prim_repo}, and we extend the implementation by adding to the routines the option of using the entropy instead of the energy during primitive recovery.

Performing an EOS table inversion with the entropy yields far smaller temperature errors than when using the energy---particularly in regions of high densities and low temperatures---and therefore, unsurprisingly, the modified routines recover the primitive variables with far smaller errors than the original. Unfortunately, because the entropy evolution assumes that entropy is conserved (an approximation that completely fails near shocks), these new routines are also only suitable as backup routines. As the entropy backup is rarely applied, and generally applied only far from shocks, we find it quite useful.

A primitive recovery step begins with guesses $W_{\rm guess}=1$ and $T_{\rm guess}=T_{\rm atm}$ or $T_{\rm guess}=T_{\rm max}$, corresponding to the atmospheric and maximum temperatures allowed in the simulation, which exist at or within the EOS table bounds. In this paper the values \mbox{$T_{\rm atm}=0.01~\mathrm{MeV}$} and \mbox{$T_{\rm max}=90~\mathrm{MeV}$} are adopted. A density guess is obtained using \mbox{$\rho_{\rm guess} = \rhostar/\left(\sqrt\gamma W_{\rm guess}\right)$} and the electron fraction is recovered analytically from \mbox{$\ye = \yestar/\rhostar$}. All other primitives are computed from these using the EOS.

Both the Palenzuela~\etal and Newman \& Hamlin routines are iterative and result in updates to the Lorentz factor and specific internal energy (or entropy), requiring EOS table inversions to compute the temperature at every iteration. Starting with the energy version of the routines and \mbox{$T_{\rm guess}=T_{\rm atm}$}, a primitive recovery is attempted using the Palenzuela~\etal routine. A failure leads to a new attempt using the Newman \& Hamlin routine. In case this fails, the temperature guess is reset to \mbox{$T_{\rm guess}=T_{\rm max}$} and the previous steps are repeated. If all previous steps fail, the process is repeated with the entropy version of the routines. If at the end of this step primitive recovery is still unsuccessful, the point is flagged and the recovery continues for the remaining grid points.

After sweeping the numerical grid once, we loop over flagged points, look at their neighbors and check for how many of them the primitive recovery has succeeded. If not enough neighbors are found the run is terminated, as clusters of failures indicate serious problems in the evolution. When the number of neighbors is sufficient, the conservative variables at the flagged points are set to
\begin{equation}
\bm{C}_{\rm new} = (1-w)\bm{C}_{\rm flagged} + w\bar{\bm{C}}_{\rm neighbors}\;,
\end{equation}
where $\bar{\bm{C}}_{\rm neighbors}$ denotes the average of the conservative variables at the neighboring points for which primitive recovery succeeded. This is repeated up to four times, successfully increasing the weight $w$ to $\sfrac{1}{4}$, $\sfrac{1}{2}$, $\sfrac{3}{4}$, and $1$ in each new attempt. If primitive recovery is unsuccessful at this point, they are reset to their atmosphere values ($\rhob=\rho_{\rm atm}$, $\ye=\ye^{\rm atm}$, and $v^{i}=0$).

We note that recovery failures are quite rare---particularly those in which all of the backup techniques fail---and typically occur in dynamically irrelevant regions: in the low density atmosphere or deep inside BH horizons. However, because production-quality simulations involve ${\sim}\rep{10^12}{10^{12}}$ primitive recovery attempts, occasional failures are a near certainty. In the case of failures, adjusting the fluid 3-velocity components to zero is rather undesirable as it could greatly and discontinuously influence the magnetic field dynamics in regions that are magnetically dominated. With this in mind, we have dedicated considerable effort in making our backup strategies robust, avoiding velocity resets as much as possible.

Finally, we note that we have not extended the Font~\etal routine to work with tabulated EOS, but this will be done in a future work. We also plan on using the tabulated EOS version of \texttt{RePrimAnd}~\cite{Kastaun:2020uxr} once it becomes available.

\subsection{Computation of optical depth}
\label{sec:computation_of_optical_depths}

In \nrpyleakage we consider the following reactions contribute to the total transport opacities $\kappa^{\nui}_{{\rm t},j}$:
\begin{alignat}{7}
  & n\medspace &&+\medspace && \nue  && \to\thickspace && \ae && +\medspace && p\;,\\
  & p\medspace &&+\medspace && \anue && \to\thickspace && \ae && +\medspace && n\;,\\
  & n\medspace &&+\medspace && \nui  && \to\thickspace && n   && +\medspace && \nui\;,\\
  & p\medspace &&+\medspace && \nui  && \to\thickspace && p   && +\medspace && \nui\;,
\end{alignat}
\rep{Explicit formulas for the neutrino emission and cooling rates appearing in Eqs.\,(\ref{eq:total_nui_emission_rate}) and (\ref{eq:total_nui_cooling_rate}), as well as for the neutrino opacities, can be found in Appendix B of~\cite{Ruffert:1995fs} and in~\cite{Burrows:2004vq}.}{%
The first two of these reactions are inverse $\beta$-processes---$\nue$ absorption onto neutrons and $\anue$ absorption onto protons, respectively---and are the dominant contributions to the optical depths of $\nue$ and $\anue$. The last two reactions describe neutral-current-scattering off neutrons, which provide the dominant contribution to the optical depths of heavy-lepton neutrinos. Explicit formulae for obtaining $\kappa^{\nui}_{{\rm t},j}$ can be found in Appendix~A of~\cite{Ruffert:1995fs}.}

Computing the local neutrino optical depths $\tau^{\nui}_{j}$ generally involves a global integration of $\kappa^{\nui}_{{\rm t},j}$ along some path $\mathcal{P}$, i.e.,
\begin{equation}
  \tau^{\nui}_{j} = \int_{\mathcal{P}}ds\,\kappa^{\nui}_{{\rm t},j}\;.\label{eq:optical_depth_integral}
\end{equation}
One common option, implemented in the open-source code \zelmanileak, is to integrate along radial rays (see e.g.,~\cite{OConnor:2009iuz,Ott:2012kr,OConnor:2011pxx}), which for simulations that use Cartesian coordinates require an auxiliary spherical grid. One interpolates data to the spherical grid, computes the opacities, and then performs the integration. This approach is particularly well-suited to study core collapse and other nearly spherically symmetric systems. However, for systems far from spherical symmetry, such as BNS and BH accretion disks, computing the optical depths this way can be very inefficient, as the resolution of the auxiliary spherical grids would need to be increased tremendously to produce accurate optical depths.

In order to have an algorithm that is both efficient and more generally applicable, we instead compute the optical depths with the local approach proposed in~\cite{Neilsen:2014hha} (see also~\cite{Siegel:2017jug,Murguia-Berthier:2021tnt}). The optical depths are first integrated to all of their nearest neighbors and then updated using the results that lead to the smallest optical depths, i.e.,
\begin{equation}
  \tau^{\nui}_{j} = \min_{\rm neighbors}\left(\tau^{\nui}_{j} + \sqrt{\gamma_{mn}\dx^{m}\dx^{n}}\kappa^{\nui}_{{\rm t},j}\right)\;,\label{eq:path_of_least_resitance}
\end{equation}
where $\dx^{i}$ is the grid spacing along the $i$\textsuperscript{th}-direction. For simplicity, we do not integrate diagonally. In this way neutrinos are allowed to explore many possible paths out of regions with relatively high optical thickness, following the path of least resistance. In our implementation, we assume that the outer boundary of the computational domain is transparent to neutrinos and thus have zero optical depths.

Because the opacities themselves depend on the optical depths, \add{the following iterative approach is used} to compute the {\it initial optical depths} at all gridpoints in the simulation domain\rem{, we implement the following iterative approach}. First the optical depths are initialized to zero, leading to an initial estimate of the opacities. This initial estimate is used to update the optical depths according to \eqref{eq:path_of_least_resitance}, which in turn allows us to recompute the opacities, and so on. One might also interpret this algorithm as considering only nearest neighbors in the first iteration, next-to-nearest neighbors in the second, next-to-next-to-nearest neighbors in the third, etc. In this way our algorithm enables neutrinos to map out paths of least resistance through arbitrary media.

When the grid structure contains multiple refinement levels, we adopt a multi-grid ``V" cycle at each iteration: the optical depths are computed on a given refinement level and, unless we are at the finest one, the solution is prolongated to the next finer one; we move to the next finer refinement level and repeat the previous step; once the finest refinement level is reached the solution is restricted to the coarser levels, completing one iteration.

The algorithm is stopped once an equilibrium is reached, measured by the overall relative change in the optical depths between consecutive iterations $(n,n+1)$,
\begin{equation}
  E = \left[\sum_{j}\sum_{\nui}\sum_{\rm interior}\left(\frac{\tau^{\nui}_{j,n+1}-\tau^{\nui}_{j,n}}{\tau^{\nui}_{j,n+1}}\right)^{2}\right]^{1/2}\;.
\end{equation}
Once $E$ falls below a user-specified threshold (typically $10^{-8}$) or the prescribed maximum allowed number of iterations (typically $2048$) is exceeded, the algorithm stops.


\section{Results}
\label{sec:results}

This section presents stress tests of these new algorithms implemented in \igm, with the aim of demonstrating both the correctness of our implementation as well as its robustness when modeling challenging astrophysical scenarios.

To this end, we first present results from challenging unit tests of the conservative-to-primitive infrastructure and the optical depth framework. Further, to validate \nrpyleakage, we evolve a simple optically thin gas and compare results between \igm and a trusted code with a physically identical but independently developed neutrino leakage scheme, \harmnuc.
Next to demonstrate the reliability of our new tabulated EOS implementation, \nrpyeos, we evolve isolated, unmagnetized NSs without neutrino leakage at different grid resolutions with different EOS tables. We consider the test as passed if numerically driven oscillations converge to zero with increased resolution at an order consistent with the reconstruction scheme (between second and third order for PPM). 

The remaining tests focus on full-scale simulations of physical scenarios that lie at the heart of multimessenger astrophysics. We simulate magnetized, equal-mass BNS systems that lead to a remnant BH with an accretion disk that is evolved for several dynamical timescales. We perform simulations with and without our new neutrino leakage scheme, comparing the qualitative differences between them. Stably modeling these last two systems, in particular, is extremely difficult if the new algorithms are not implemented correctly, thus acting as the most strenuous tests conducted in our study.

In tests involving tabulated EOSs, we use three different fully-tabulated microphysical EOSs: the Lattimer--Swesty EOS with incompressibility modulus $K=220~\mathrm{MeV}$~\cite{Lattimer:1991nc} (henceforth LS220), the Steiner--Hempel--Fisher EOS~\cite{Steiner:2012rk} (henceforth SFHo), and the SLy4 EOS of~\cite{Chabanat:1997un}. For the first three, we use the tables by O'Connor--Ott~\cite{OConnor:2009iuz}, while for the last one we use the table by Schneider--Roberts--Ott~\cite{Schneider:2017tfi}, all of which are freely available at~\cite{stellarcollapse_website}. These choices of EOS were made largely to facilitate direct comparisons with \harmnuc in the case of BH accretion disks.

Finally, we note that the EOS tables were cleaned with a simple script to change the reference mass used in the tables in such a way that the specific internal energy is never negative. We also clean up some of the table entries to avoid superluminal sound speeds. These are standard procedures when using EOS tables from~\cite{stellarcollapse_website}.

\subsection{Primitive variables recovery}
\label{sec:primitive_variables_recovery}

In order to validate our implementation of primitive recovery routines described in \secref{sec:conservative_to_primitive}, we perform similar tests to the ones described in~\cite{Siegel:2017sav,Murguia-Berthier:2021tnt}. First a set of primitive variables is specified and conservatives computed. Then the primitive recovery routine is selected and the conservatives injected into it. The primitives that are output from the recovery routine are compared with the input primitives, producing an estimate of how well the tested routine is able to recover the correct primitive variables. We refer to this as a ``$\bm{P}$ to $\bm{C}$ to $\bm{P}$'' test.

We measure the primitive recovery error at each recovery as a sum of relative errors across all primitive variables in the primitives vector $\bm{P}$
\begin{equation}
  E_{\bm{P}\to\bm{C}\to\bm{P}} = \sum_{i}\left|1-\frac{p^{\rm recovered}_{i}}{p^{\rm original}_{i}}\right|\;,\label{eq:EPtoCtoP}
\end{equation}
where $p^{\rm original}_{i}$ and $p^{\rm recovered}_{i}$ represent the original and recovered primitive variables, respectively, and the sum includes all variables in \eqref{eq:prims}.

For a given EOS, \add{we perform two separate tests in which} we use \mbox{$N=2^{12}$} points to evenly discretize \rep{$\log_{10}\rhob\in\left[-12,-3\right]$ and $\log_{10}T\in\left[-2,2\right]$}{non-constant hydrodynamic quantities}. \rep{We}{In the first test we consider $\log_{10}\rhob\in\left[-12,-3\right]$,  $\log_{10}T\in\left[-2,2\right]$ and} arbitrarily fix the electron fraction to \mbox{$\ye=0.1$}\add{, the Lorentz factor to \mbox{$W=2$}, and the magnetic pressure to \mbox{$\log_{10}\bigl(P_{\rm mag}/P\bigr)=-5$}; in the second we consider $\log_{10}\bigl(P_{\rm mag}/P\bigr)\in\left[-5,9\right]$, $\log_{10}\bigl(W-1\bigr)\in\left[-5.5,1.5\right]$, and arbitrarily fix \mbox{$\rhob=10^{11}\;\mathrm{g/cm^{3}}$}, \mbox{$Y_{\rm e}=0.1$}, and \mbox{$T=5\;\mathrm{MeV}$}.} \rep{and compute t}{T}he pressure, specific internal energy, and entropy \add{are computed} using the EOS table\rep{. T}{, while t}he spatial components of the velocities and magnetic fields are set randomly \rep{assuming $W=2$ and $\log_{10}(P_{\rm mag}/P)=-5$}{from $W$ and $P_{\rm mag}$}, respectively. Finally, we compute the conservative variables using \eqref{eq:conservs} assuming flat space.

\begin{figure}[htb!]
  \centering
  \includegraphics*[width=\columnwidth]{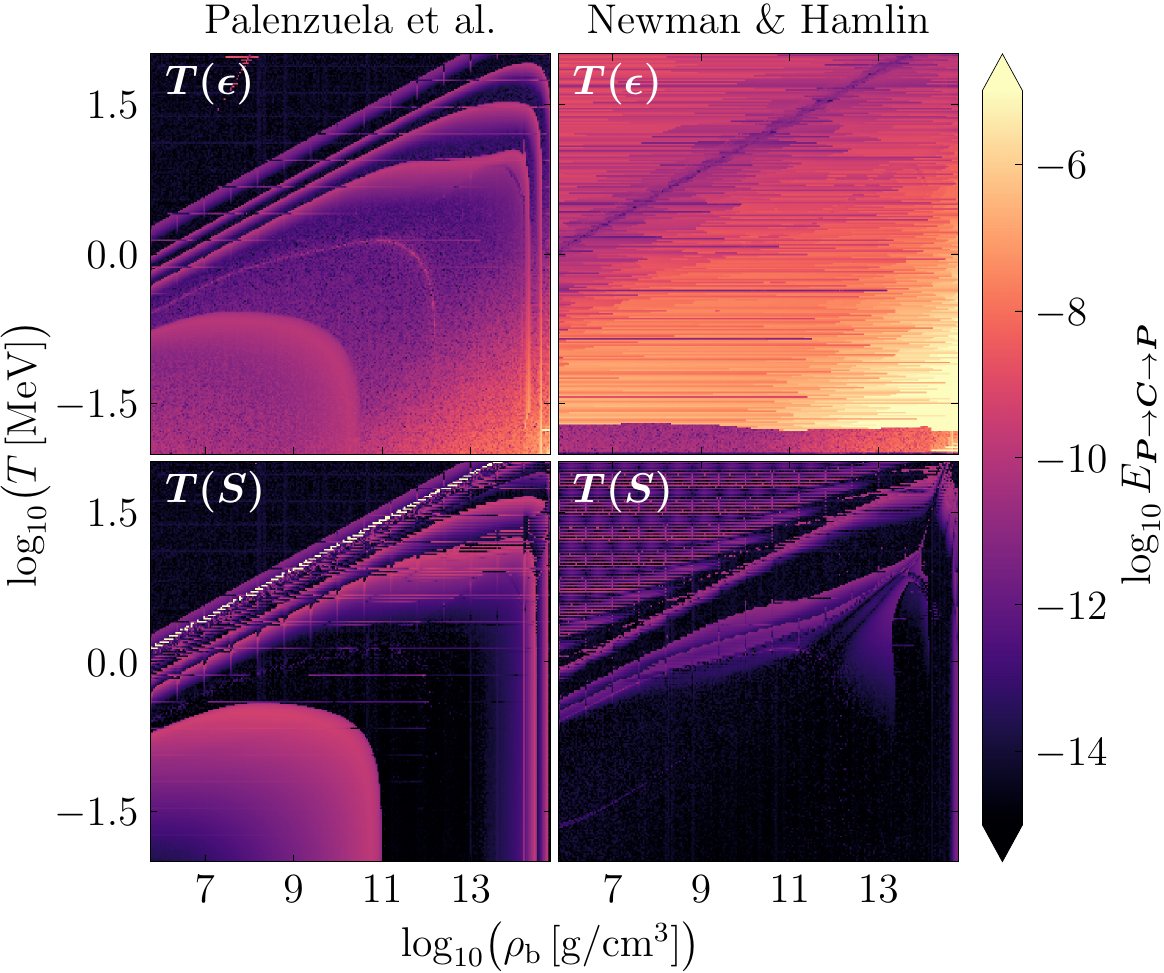}\\[0.2cm]
  \includegraphics*[width=\columnwidth]{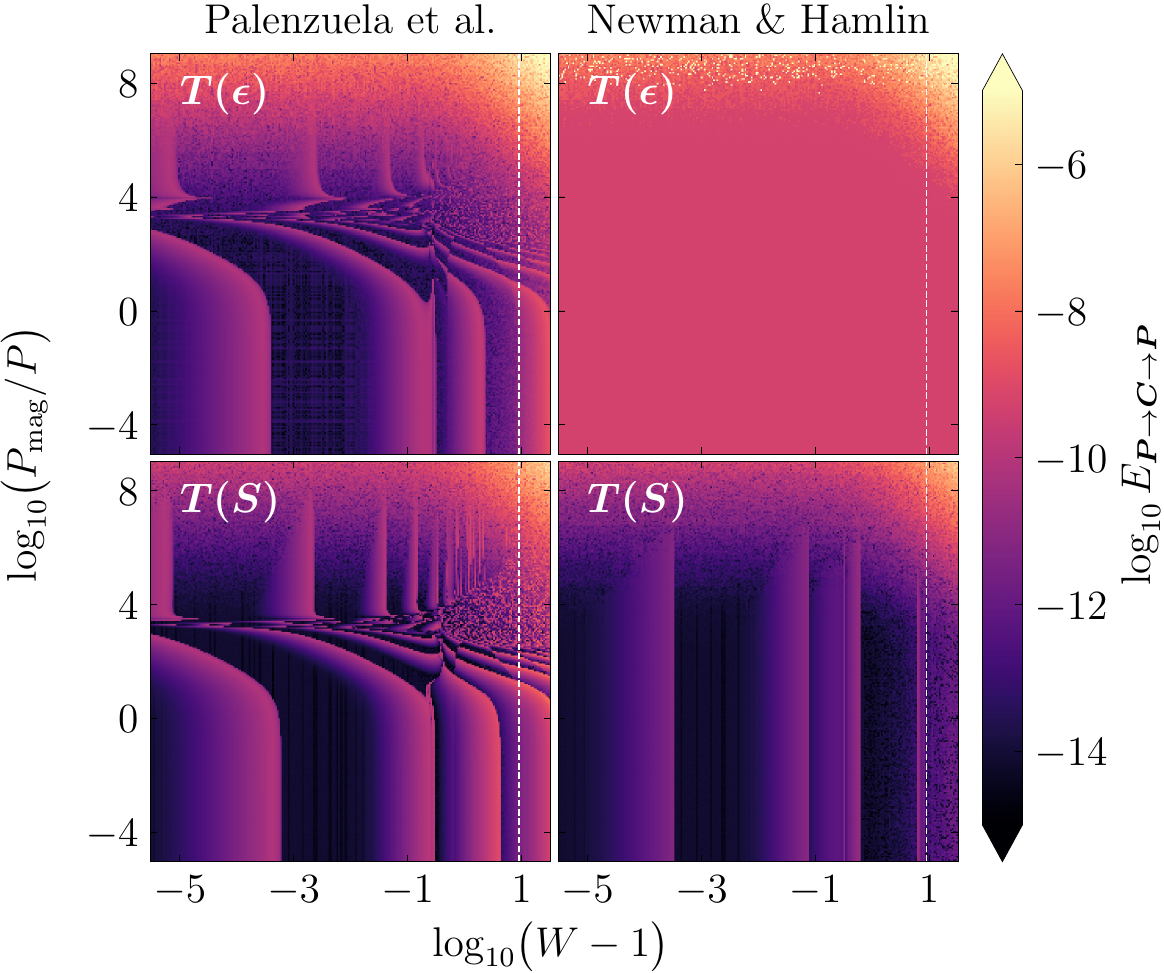}
  \caption{Primitive recovery errors (\eqrefalt{eq:EPtoCtoP}) from \mbox{``$\bm{P}$ to $\bm{C}$ to $\bm{P}$''} tests using the Palenzuela \etal and Newman \& Hamlin routines, as implemented in \igm. \textbf{Top}/\textbf{bottom}: \add{Tests performed with fixed \mbox{$\bigl(\ye,P_{\rm mag},W\bigr)\big/\bigl(\rhob,\ye,T\bigr)$}. In each figure's top/bottom panels,} table inversions to obtain the temperature are performed using the specific internal energy/entropy. \add{The vertical dashed white lines in to bottom figure indicate \mbox{$W=10$}, the maximum Lorentz factor allowed in our simulations.}
  }
  \label{fig:con2prim}
\end{figure}

We use this setup to test the Palenzuela \etal and the Newman \& Hamlin routines as implemented in \igm. As described in \secref{sec:conservative_to_primitive}, each routine is given two chances to recover the primitives, with initial guesses $T_{\rm guess}=T_{\rm atm}$ and, in case of failure, $T_{\rm guess}=T_{\rm max}$. In \figref{fig:con2prim} we present the test results.

As our implementation allows for one to use either the specific internal energy or the entropy to recover the temperature, we perform tests using both. Our results indicate that using the entropy generally leads to errors that are at least comparable to, but often smaller than, those obtained using the energy. However, because \igm currently adopts an approximate entropy evolution equation (assuming entropy is conserved), we are unable to reliably use the entropy as our default variable for primitive recovery. During actual numerical evolutions, we thus limit use of the entropy variable to a backup when recovery using the specific internal energy fails.

\add{Further stress tests of the routines indicate they are robust for magnetizations as high as \mbox{$\log_{10}\bigl(P_{\rm mag}/P\bigr)\sim15$}, although the recovery errors increase for \mbox{$\log_{10}\bigl(P_{\rm mag}/P\bigr)\gtrsim10$}. We also found that the routines typically fail to recover the primitives for \mbox{$W\gtrsim100$}, which is not worrisome given we impose a hard ceiling on the Lorentz factor which is much smaller than this value (\mbox{$W_{\rm max}=10$} in the simulations presented in this paper).}

\subsection{Optically thin gas}
\label{sec:optically_thin_gas}

To validate our neutrino leakage implementation \nrpyleakage, we model an isotropic gas of constant density at rest in flat space assuming the SLy4 EOS. \add{We note that this system has been chosen for its simplicity, not its physical relevance, as it allows us to validate the implementation of our leakage scheme against a trusted code.}

In this scenario, the GRMHD equations simplify to
\begin{equation}
    \partial_{t}\ye = \RR/\rhob\quad\text{and}\quad\partial_{t}\epsilon = \QQ/\rhob\;.
    \label{eq:optically_thin_simple}
\end{equation}
These equations are solved straightforwardly with a standalone code, which supports both \nrpyleakage and the leakage scheme of \harmnuc. We then compare the results from these equations against those generated when \igm evolves the full set of GRMHD equations. Agreement between \igm and the standalone code provides an external validation of \nrpyleakage in the optically thin regime, as well as its integration within \igm.

The solution has the following behavior. When the electron fraction is large (small), electron (positron) capture by protons (neutrons) is favored, resulting in a decrease (increase) of electron neutrinos (antineutrinos) and a decrease (increase) of the electron fraction with time. Note that, by construction, $\QQ\leq0$ and thus the specific internal energy and temperature are always expected to decrease.

We perform two tests to verify the expected behavior of the system as described above. In one test we set the initial electron fraction to \mbox{$\ye(0)=0.5$}, while in the other we set it to \mbox{$\ye(0)=0.005$}. In both cases the density and initial temperature of the gas are set to \mbox{$\rhob=10^{-12}$} and \mbox{$T(0)=1~\mathrm{MeV}$}, respectively. Other hydrodynamic quantities, like the initial specific internal energy, are computed as needed using the SLy4 EOS of~\cite{Schneider:2017tfi}.

\begin{figure}[htb!]
    \centering
    \includegraphics[width=\columnwidth,clip]{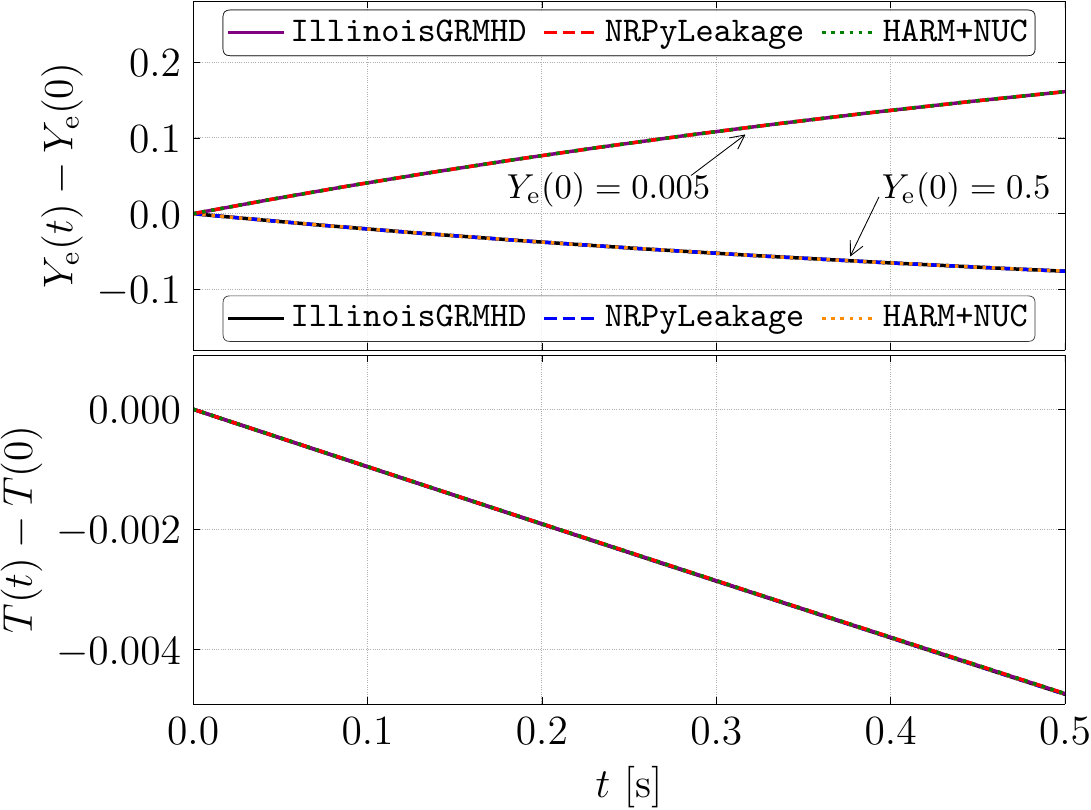}
    \caption{Electron fraction (\textbf{top}) and temperature (\textbf{bottom}) evolution for an isotropic gas of constant density, for different initial values of $\ye$.}
    \label{fig:isotropic_gas_results}
\end{figure}

In \figref{fig:isotropic_gas_results} we show the excellent agreement between the results obtained by the different codes.%
\footnote{The leakage scheme in \harmnuc assumes that the optical depths are always large when computing the neutrino degeneracy parameters. This is a reasonable assumption, given that most physical systems of interest are not transparent to neutrinos. Nevertheless, this causes a discrepancy between the leakage scheme in \harmnuc and \nrpyleakage. For the sake of the comparison made here, we slightly modified the way \harmnuc computes the neutrino degeneracy parameters to match what is done in \nrpyleakage; i.e., using Eqs.\,(A3) and (A4) in~\cite{Ruffert:1995fs}.}

\subsection{Optically thick sphere}
\label{sec:optically_thick_sphere}

We next demonstrate the robustness of our implementation of the optical depth initialization algorithm on Cartesian AMR grids. To this end, a simple case of a sphere of constant density, electron fraction, and temperature is considered, \add{for} which \rem{implies that} the opacities \rep{in the sphere are also constant and therefore the optical depths can be determined analytically with \eqref{eq:optical_depth_integral}}{vanish everywhere except in the interior of the sphere}.

\add{Given the opacities for this spherically symmetric system, the optical depths can also be determined semi-analytically by performing the integral \eqref{eq:optical_depth_integral} along radial rays, allowing us to validate our iterative algorithm. Since \mbox{$\left(\rhob,\ye,T\right)$} are constant, so are the opacities, yielding
\begin{equation}
    \tau_{j}^{\nui} = 
    \left\{
    \begin{matrix}
        \kappa_{j}^{{\rm t},\nui}(r_{\rm Sph}-r) &,& r\in[0,r_{\rm Sph}] &,\\
        0 &,& \text{otherwise} &.
    \end{matrix}
    \right.
    \label{eq:tau_semianalytic}
\end{equation}
}

\rep{Specifically, t}{T}he sphere is assumed to have radius \mbox{$r_{\rm Sph}=2.5$}, \rem{and} constant density \mbox{$\rhob^{\rm Sph}=9.8\times10^{13}~\mathrm{g/cm^{3}}$}, electron fracton \mbox{$\ye^{\rm Sph}=0.1$}, and temperature \mbox{$T^{\rm Sph}=8.0~\mathrm{MeV}$} in an optically thin medium with \mbox{$\rhob^{\rm Ext}=6\times10^{7}~\mathrm{g/cm^{3}}$}, \mbox{$\ye^{\rm Ext}=0.5$}, and \mbox{$T^{\rm Ext}=0.01~\mathrm{MeV}$}. We adopt the SLy4 EOS of~\cite{Schneider:2017tfi}.

\begin{figure}[htb!]
  \centering
  \includegraphics[width=\columnwidth]{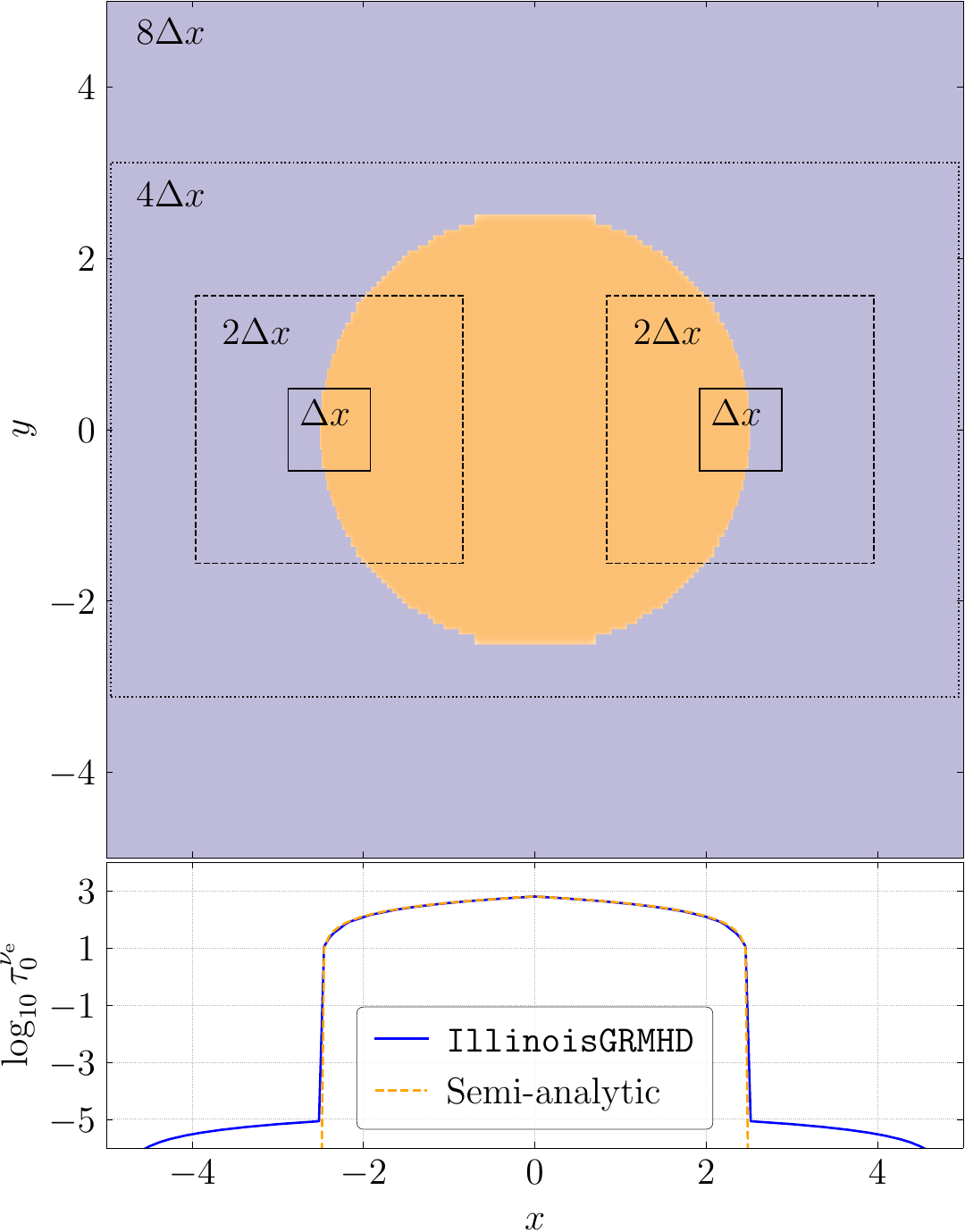}
  \caption{Stress test for the ``path of least resistance'' algorithm used to compute the optical depths. \textbf{Top}: Our grid setup and the constant density sphere. \textbf{Bottom}: Comparison between the results obtained using \igm and the exact solution along $y=0$.}
  \label{fig:const_dens_sphere_results}
\end{figure}

Our grid is a Cartesian box of side-length $10r_{\rm Sph}$ with four refinement levels, as illustrated in the upper panel of \figref{fig:const_dens_sphere_results}. We add two refinement centers located at ${\pm}2.5$, each with three levels of refinement. Of course, this grid structure would never be used to simulate a spherical object, but because the surface of the sphere crosses multiple refinement boundaries, it provides a significantly challenging test for our optical depth initialization algorithm as detailed in \secref{sec:computation_of_optical_depths}. We find excellent agreement with the \rep{exact}{semi-analytic} results \add{obtained using the opacities from the iterative algorithm in \eqref{eq:tau_semianalytic}}, as shown in the bottom panel of the figure.

\subsection{Tolman--Oppenheimer--Volkoff star}
\label{sec:tov_star}

As our next validation test, we evolve unmagnetized, stable Tolman--Oppenheimer--Volkoff (TOV) NSs. In this test, we disable neutrino leakage to ensure the NS maintains this equilibrium solution in the continuum limit. That is to say, in the limit of infinite numerical resolution, we expect zero oscillations in our simulated NSs.

When placing the stars in our finite resolution numerical grids however, numerical errors induce stellar oscillations. Because these oscillations are largely caused by truncation error associated with \igm's reconstruction scheme, they should converge to zero as we increase the resolution of the numerical grid. To confirm this, we perform simulations at three different resolutions---hereafter low (LR), medium (MR), and high (HR)---and demonstrate that the oscillations converge away at the expected rate.

To obtain initial data, a tabulated EOS is chosen and the TOV equations are solved using \nrpy's TOV solver. Three different EOS tables are used: LS220, SFHo, and SLy4. The initial temperature is fixed to \mbox{$T=0.01~\text{MeV}$}, while the initial electron fraction is determined by imposing the neutrino-free beta-equilibrium condition \mbox{$\mu_{\nu}(\rhob,\ye,T)=0$}, where $\mu_{\nu}$ is the neutrino chemical potential. The initial data are then evolved forward in time with the \etk, using \baikal~\cite{nrpytutorial} and \igm to perform the spacetime and GRMHD evolutions, respectively.

\begin{figure}[htb!]
  \centering
  \includegraphics[width=\columnwidth,clip]{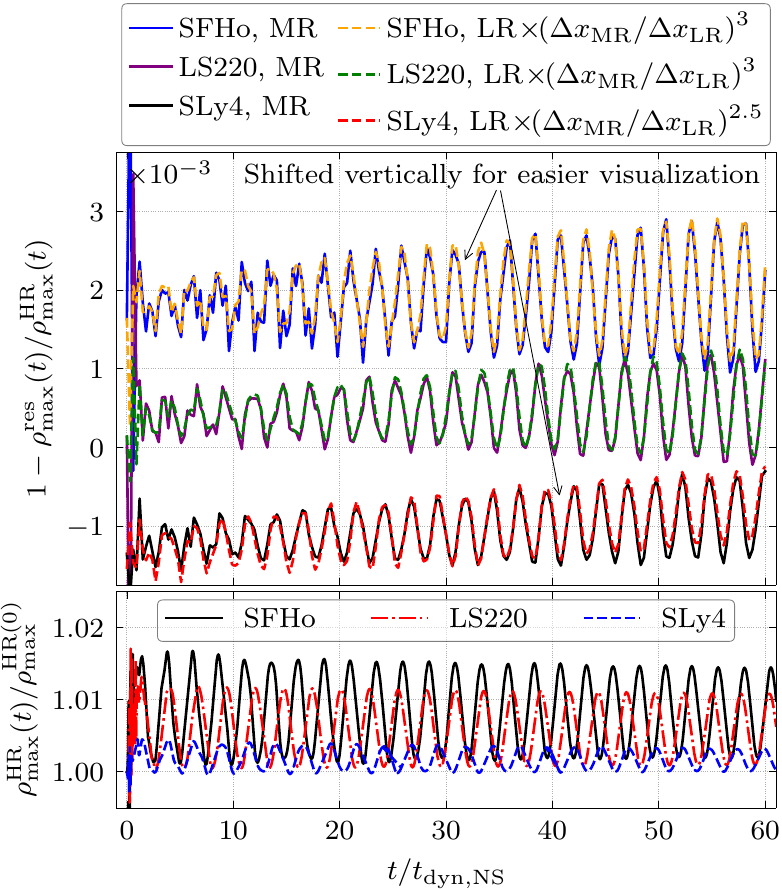}
  \caption{\textbf{Bottom}: High resolution (HR) evolution of the maximum density for unmagnetized TOV stars of $M=1.4\Msun$ and different EOSs. \textbf{Top}: Relative errors in density oscillations between the HR run and the low (LR) and the medium (MR) resolutions runs. Here \mbox{$t_{\rm dyn,NS}^{\rm SFHo}\approx0.134~\mathrm{ms}$}, \mbox{$t_{\rm dyn,NS}^{\rm LS220}\approx0.144~\mathrm{ms}$}, and \mbox{$t_{\rm dyn,NS}^{\rm SLy4}\approx0.131~\mathrm{ms}$} (\eqrefalt{eq:t_dyn}).}
  \label{fig:tov_noleak_density}
\end{figure}

The numerical grid structure contains five factor-of-two levels of Cartesian AMR, with resolutions on the finest level of \mbox{$\dx_{\rm LR}=1.5\dx_{\rm MR}=2\dx_{\rm HR}\approx277~\mathrm{m}$}. The stars are evolved for approximately $60$ dynamical timescales
\begin{equation}
  t_{\rm dyn,NS} = \frac{1}{\sqrt{\rho_{0,{\rm max}}}}\;,\label{eq:t_dyn}
\end{equation}
where $\rho_{0,{\rm max}}$ is the maximum initial density (i.e., the density at the center of the NS). We monitor the maximum density on the grid as a function of time, with results from the HR runs displayed on the bottom panel of \figref{fig:tov_noleak_density}.

The top panel of \figref{fig:tov_noleak_density} displays the relative errors, $E$, of the oscillations from the MR and LR runs against the oscillations from the HR run. Assuming $E\propto\dx^{p}$, we find
\begin{equation}
  E_{\rm MR} = \left(\frac{\dx_{\rm MR}}{\dx_{\rm LR}}\right)^{p}E_{\rm LR}\;.
\end{equation}
Thus, for a numerical scheme that is $p$-order accurate, we expect that multiplying the relative errors of the LR run by $(\dx_{\rm MR}/\dx_{\rm LR})^{p}$ will yield similar errors as the MR run. Generally we would expect the numerical errors to be dominated by our reconstruction method (PPM), which is between second and third order. This is indeed the observed behavior---the convergence order of our numerical scheme is found to be $p\in[2.5,3]$.

\subsection{Magnetized binary neutron stars}
\label{sec:bns}

With the core new features validated, we now turn our attention to fully dynamical GRMHD simulations of a magnetized, equal-mass BNS system, modeling the inspiral, merger, and the resulting remnant BH. We adopt the LS220 microphysical, finite-temperature EOS, and simulate the system both with and without neutrino leakage enabled. This self-validation test is quite challenging, and is bound to result in unphysical behavior or, in the worst case, code crashes, if there are issues in our implementation.

Initial data are obtained using \lorene~\cite{Gourgoulhon:2000nn,Feo:2016cbs,2016ascl.soft08018G,lorene_website}. The initial separation of the system is $45~\mathrm{km}$ and the gravitational (baryonic) mass of each NS is \mbox{$1.39 M_{\odot}$} (\mbox{$1.59 M_{\odot}$}), while the total ADM mass of the system is \mbox{$M_{\rm ADM}= 2.86 M_{\odot}$}. The interior of each NS is seeded with a poloidal magnetic field with maximum initial value ${\sim}10^{15}~G$ (see Appendix C of~\cite{Etienne:2015cea}). The initial temperature is set to $0.01~\text{MeV}$ and the electron fraction is determined by imposing the neutrino-free beta-equilibrium condition.

\baikal, which solves Einstein's equations in the BSSN formalism, is used to evolve the spacetime. Again the \carpet AMR infrastructure is adopted to set up a grid with eight refinement levels by factors of two, with the resolution at the finest level \mbox{$\dx_{8}\approx185~\mathrm{m}$}. Upon BH formation, two additional refinement levels are added to better resolve the puncture, and thus the highest resolution on the grid becomes \mbox{$\dx_{10}\approx46~\mathrm{m}$}.

Performing the simulation using the \etk gives us access to outstanding diagnostic thorns, of which prominent examples include \ahfd~\cite{Thornburg:2003sf}---used to locate and compute the shape of apparent horizons---and \qlm~\cite{Dreyer:2002mx,Schnetter:2006yt}---used to compute useful quasi-local quantities like the BH mass and spin. Additionally, \igm carefully monitors the number of times the primitive recovery infrastructure resorts to backup strategies, as well as the strategies used, aborting the simulation if any major error is detected. For the two simulations performed in this paper, we have not observed atmosphere resets or conservative averages (see \secref{sec:conservative_to_primitive}), which reflects the robustness of our conservative-to-primitive implementation.

\begin{figure*}[htb!]
  \centering
  \includegraphics[width=0.7\textwidth]{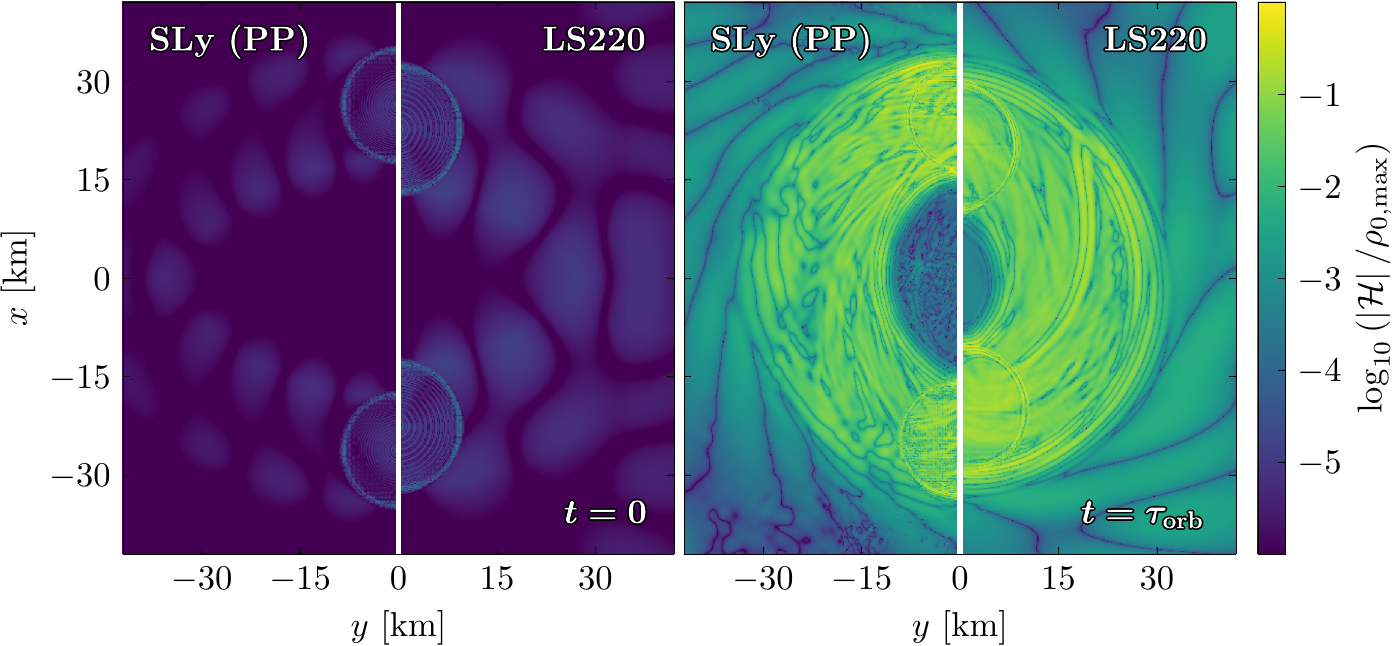}
  \caption{Hamiltonian constraint violations normalized by the initial maximum baryonic density, comparing the PPEOS SLy against the LS220 tabulated EOS in the context of an equal-mass magnetized BNS simulation. Results are plotted on the orbital plane. {\bf Left}: initial constraint violations; {\bf right}: constraint violations after one full orbit.}
  \label{fig:bns_hamiltonian_constraint_comparison}
\end{figure*}

Satisfaction of the Einstein constraints reflect the health of BNS simulations in a holistic sense. As evidence that our updated implementation of \igm is working correctly, we carefully monitor the Hamiltonian constraint violation throughout the numerical evolution. In Fig.~\ref{fig:bns_hamiltonian_constraint_comparison} the magnitude of these violations with the LS220 tabulated EOS is compared against an evolution of an SLy piecewise polytropic EOS (PPEOS) BNS evolution performed with a trusted version of \igm with hybrid equation of state support (adopting a PPEOS for the cold pressure). Both are equal-mass binaries (but run without any symmetries imposed), with each neutron star having a baryonic mass of $1.49\Msun$ and $1.59\Msun$ in the tabulated and PPEOS cases, respectively. The initial separations are different, so as to reuse existing data; this has no bearing on our assessment. 
The left panel of \figref{fig:bns_hamiltonian_constraint_comparison} displays side-by-side comparisons of this diagnostic at $t=0$ and, in the right panel, after a full orbit at $t=\tau_{\rm orb}$.%
\footnote{\add{Note that $\tau_{\rm orb}$ corresponds to different physical times in different simulations, but is unambiguously defined as the time it takes the stars to complete their first orbit.}}
Initial violations are smaller for the PPEOS case, as the \lorene initial data possessed higher spectral resolution. The small initial constraint violation difference is quickly dominated by numerical-evolution errors, such that after one full orbit, constraint violations quickly reach a steady-state a few orders of magnitude higher than in the initial data. Generally we would expect that errors due to EOS table interpolations would result in slightly higher constraint violations in the tabulated EOS case, but we find that both runs exhibit quite comparable results.

\begin{figure}[htb!]
  \centering
  \includegraphics[width=\columnwidth]{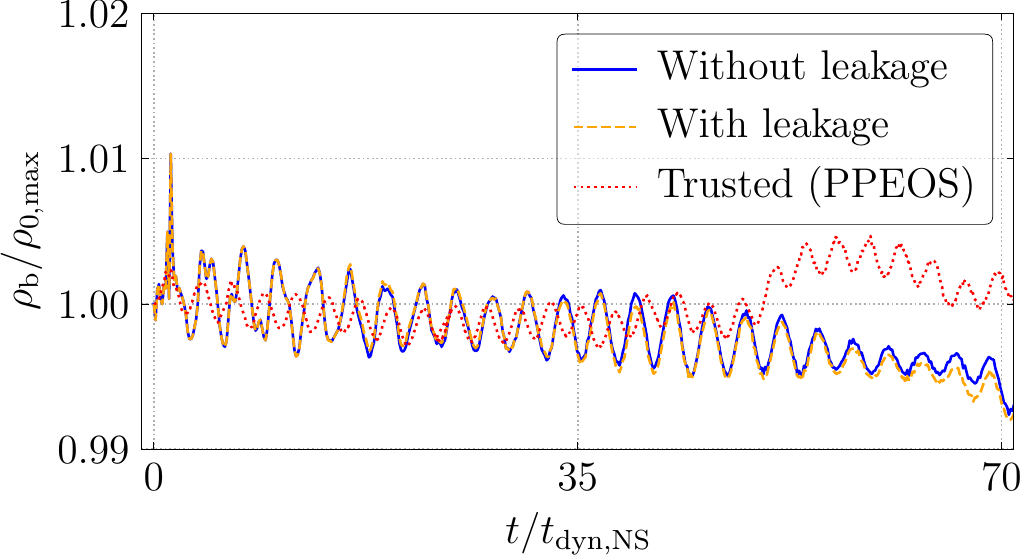}
  \caption{Maximum baryonic density as a function of time, comparing the new version of \igm, both with (orange dashed) and without (blue solid) neutrino leakage, against the previous, trusted version of the code that implements PPEOS (red dotted). Note that $t_{\rm dyn,NS}\approx0.141~\mathrm{ms}$ (\eqrefalt{eq:t_dyn}).}
  \label{fig:bns_density_comparison}
\end{figure}

As in the TOV tests, we also track the maximum density on the grid over time---i.e., the density at the center of the NSs during inspiral. \figref{fig:bns_density_comparison} shows the evolution of this quantity for 70 dynamical timescales, which happens to be shortly before merger in the tabulated EOS (LS220) case. As can be seen in the figure, the maximum density remains constant to 0.75\% of its initial value throughout, indicating that the hydrostatic equilibrium of the NSs, imposed by the initial data, is maintained through merger, and to a degree that is comparable or better when comparing the latest \igm against the trusted PPEOS version. Further, as expected, data in this figure demonstrate that neutrino leakage has no impact on the central densities of the neutron stars.

\begin{figure}[htb!]
  \centering
  \includegraphics[width=\columnwidth]{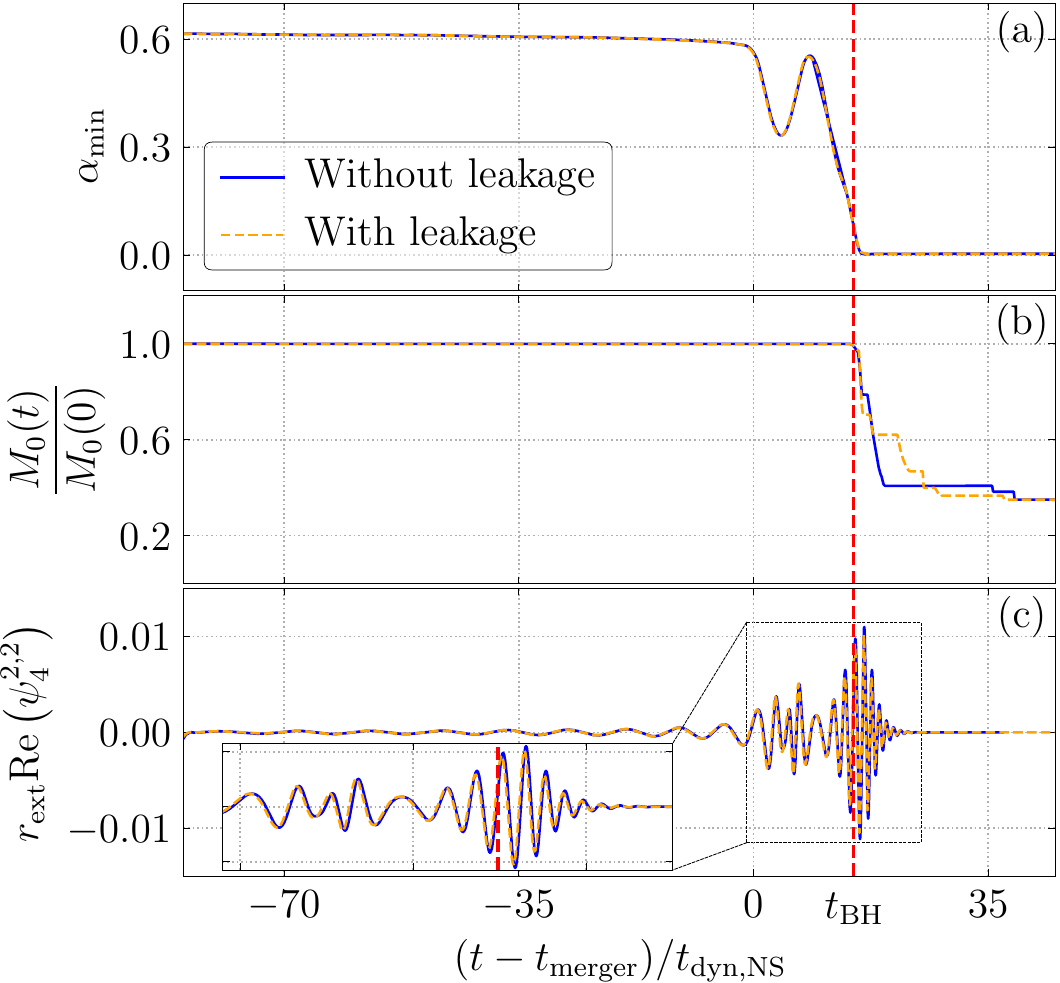}
  \caption{Several diagnostics extracted from magnetized, equal-mass BNS simulations performed with \igm using the LS220 microphysical, finite-temperature EOS. Results obtained without (with) neutrino leakage are displayed using blue, solid (orange, dashed) curves. The red, dashed line shows the time of BH formation $t_{\rm BH}$, while $t_{\rm merger}$ corresponds to the time where the stars first touch.
  $(a)$ Minimum value of the lapse function.
  $(b)$ Conservation of the rest mass of the system, computed using \eqref{eq:rest_mass}.
  $(c)$ Dominant $(2,2)$ mode of the gravitational wave strain extracted at \mbox{$r_{\rm ext}\approx738.3\ {\rm km}$}. We have subtracted the time it takes for the wave to propagate to $r_{\rm ext}$ when generating the plot. Note that \mbox{$t_{\rm dyn,NS}\approx0.141~\mathrm{ms}$}, \mbox{$t_{\rm merger}\approx12.2~\mathrm{ms}$}, and \mbox{$t_{\rm BH}\approx14.3~\mathrm{ms}$}.}
  \label{fig:bns_lapse_rest_mass_gw}
\end{figure}

Next we focus on the merger of the tabulated EOS (LS220) BNS simulations, comparing results both with and without neutrino leakage enabled. BH formation occurs in coincidence with the collapse of the lapse function toward zero. When \mbox{$\alpha_{\rm min}<0.1$}, we trigger \carpet to add additional refinement levels to our grid (as previously described) so that the moderately spinning black hole is sufficiently resolved.

Further, post-merger oscillations of \mbox{$\alpha_{\rm min}(t)$} are monitored as an indication of how close the merger remnant (a very short-lived HMNS in this case) is to BH formation.
As can be seen in \figrefalt{fig:bns_lapse_rest_mass_gw}{a}, our simulations lead to a HMNS that undergoes a single oscillation prior to collapse to a BH, with an apparent horizon detected \mbox{$t_{\rm BH}\approx2.1~\mathrm{ms}$} after merger. Comparing with a couple other equal-mass results in the literature with similar initial NS masses adopting this EOS, we find that both result in a short-lived HMNS remnant that collapses to a BH between \mbox{$t_{\rm BH}\approx8.5~\mathrm{ms}$} (\cite{Kastaun:2014fna}, with each NS having isolated ADM mass \mbox{$M_{\rm NS}=1.41\Msun$}) and \mbox{$t_{\rm BH}\approx48.5~\mathrm{ms}$} (\cite{Bernuzzi:2015opx}, with each NS having isolated ADM mass \mbox{$M_{\rm NS}=1.35\Msun$}). However a clean, apples-to-apples comparison cannot be made, as simulations in these references did not include magnetic fields, chose different initial separations, and adopted different numerical resolutions/grids. Indeed, further work is sorely needed to cleanly validate current HMNS lifetime estimates across different codes, and we plan to perform such comparisons in future work.

As further validation that our conservative GRMHD scheme is working correctly, conservation of rest mass, i.e.,
\begin{equation}
  M_{0}=\int W\rho_{\rm b}\sqrt{\gamma}\,dV\; = {\rm constant},\label{eq:rest_mass}
\end{equation}
is carefully monitored. Provided GRMHD flows do not cross AMR refinement levels or a black hole forms, this conservation should be maintained to roundoff error. Indeed, \figrefalt{fig:bns_lapse_rest_mass_gw}{b} demonstrates that the initial value is conserved to 0.001\% throughout the inspiral for over 75 dynamical timescales. During and after BH formation, a significant amount of rest mass forms and falls into the horizon, where ceilings on maximum density are imposed for numerical stability and resulting in a loss of rest mass.

Finally, the gravitational wave signal, from the inspiralling of two NSs through the oscillations of the HMNS, to the ringing BH, is one of the key theoretical predictions derived from these simulations. In \figrefalt{fig:bns_lapse_rest_mass_gw}{c} we display the real component of the dominant, $(2,2)$ mode of $\psi_{4}$ extracted at \mbox{$r_{\rm ext}\approx738.3\ {\rm km}$}. We note that in this panel of the figure we also subtract the time that it takes for the wave to propagate from the center of the grid to $r_{\rm ext}$.

Minor differences in the signal are observed between simulations with and without neutrino leakage, which we attribute to the great increase in neutrino production after the massive shock---and consequent heating---that occurs when the NSs merge. As the neutrinos leak, they carry energy and deleptonize the system, which help explain the small differences in the gravitational wave signals.

\begin{figure*}[!h]
  \centering
  \begin{tabular}{c|c}
    Neutrino leakage \textbf{enabled} & Neutrino leakage \textbf{disabled}\\
    \subfloat[$t=0~\mathrm{ms}$.]{
      \includegraphics[width=0.45\textwidth,clip]{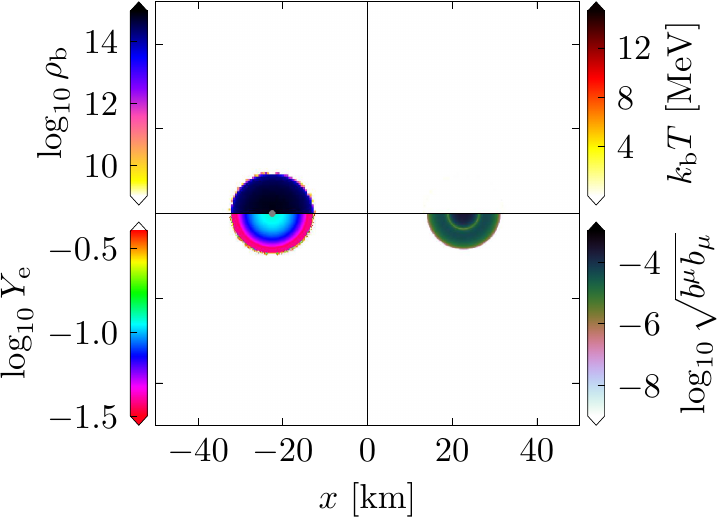}
      \label{fig:bns_leak1}
    }
    &
    \subfloat[$t=0~\mathrm{ms}$.]{
      \addtocounter{subfigure}{2}
      \includegraphics[width=0.45\textwidth,clip]{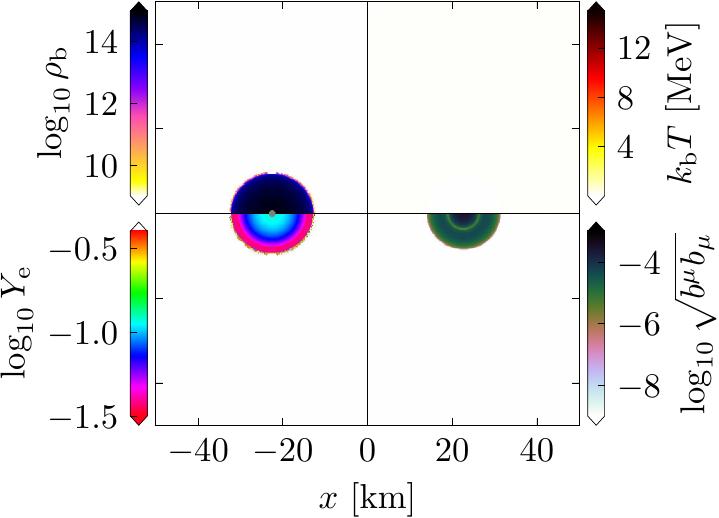}
      \label{fig:bns_noleak1}
    }
    \\
    \subfloat[$t=34t_{\rm dyn,NS}=4.8~\mathrm{ms}$.]{
      \addtocounter{subfigure}{-3}
      \includegraphics[width=0.45\textwidth,clip]{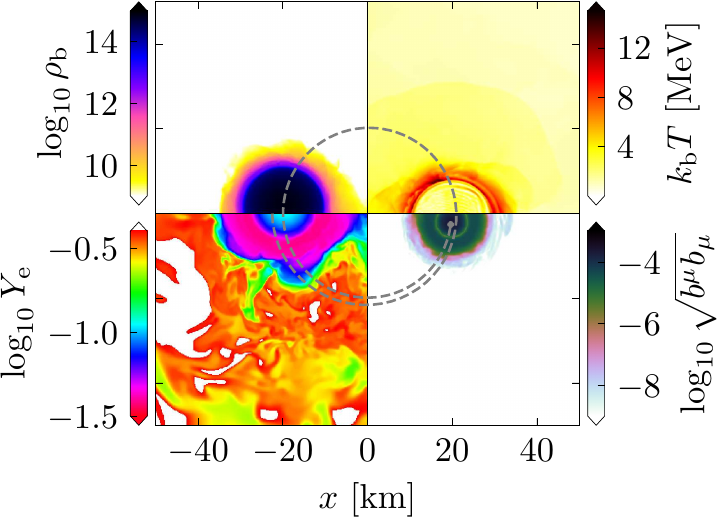}
      \label{fig:bns_leak2}
    }
    &
    \subfloat[$t=34t_{\rm dyn,NS}=4.8~\mathrm{ms}$.]{
      \addtocounter{subfigure}{2}
      \includegraphics[width=0.45\textwidth,clip]{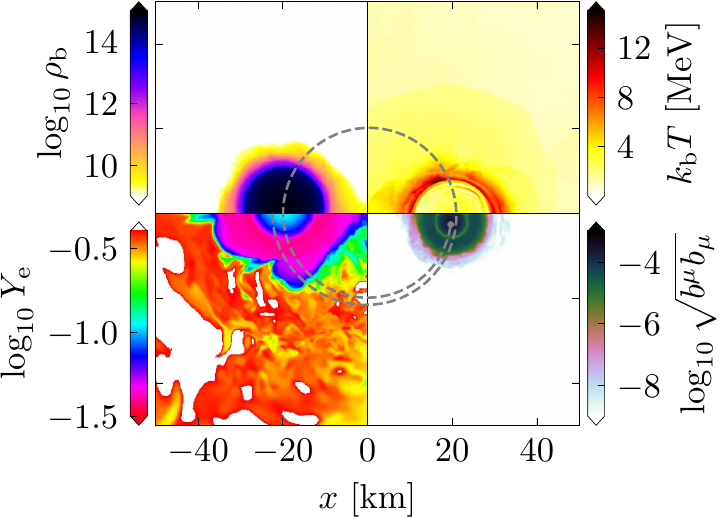}
      \label{fig:bns_noleak2}
    }
    \\
    \subfloat[$t=68t_{\rm dyn,NS}=9.6~\mathrm{ms}$.]{
      \addtocounter{subfigure}{-3}
      \includegraphics[width=0.45\textwidth,clip]{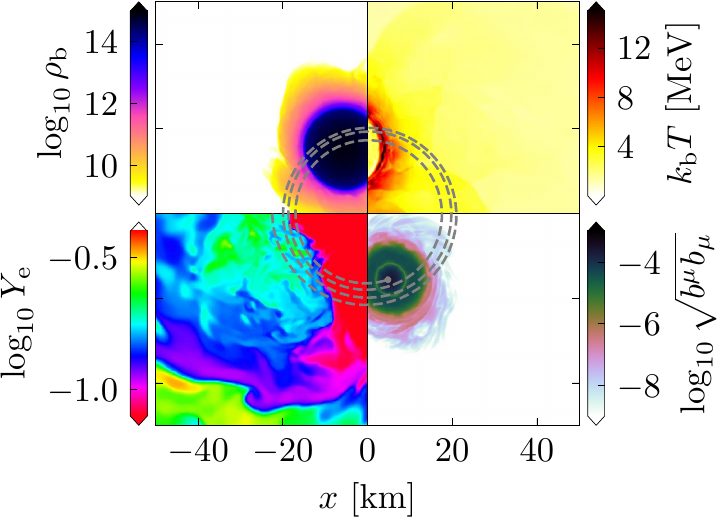}
      \label{fig:bns_leak3}
    }
    &
    \subfloat[$t=68t_{\rm dyn,NS}=9.6~\mathrm{ms}$.]{
      \addtocounter{subfigure}{2}
      \includegraphics[width=0.45\textwidth,clip]{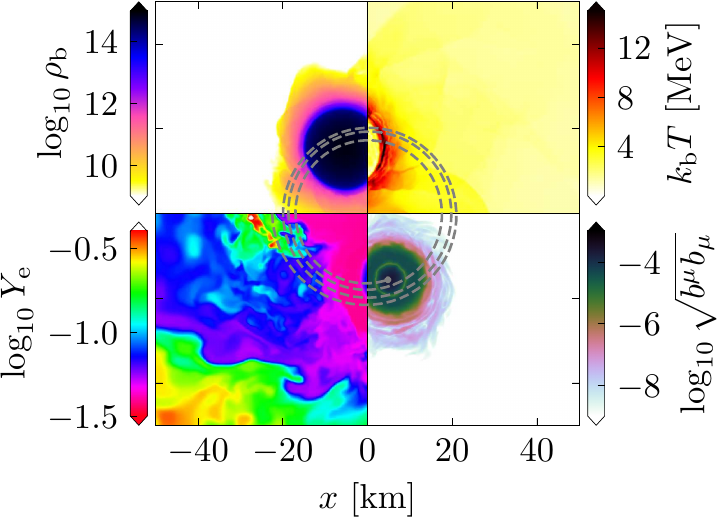}
      \label{fig:bns_noleak3}
    }
  \end{tabular}
  \caption{Time evolution of a magnetized, equal-mass BNS system using microphysical, finite-temperature EOS. Each panel shows (starting from the top left and moving clockwise) the logarithm of the density (in $\mathrm{g/cm^{3}}$), the temperature in MeV, the logarithm of $\sqrt{b^{\mu}b_{\mu}}$, and the logarithm of the electron fraction on the orbital ($xy$) plane. Panels (a)--(c) and (d)--(f) display the inspiral phase of the simulation with neutrino leakage enabled and disabled, respectively. The trajectory of the star initially on the left of the grid is shown using a gray, dashed line and its center is indicated by a gray dot. Note that \mbox{$t_{\rm dyn,NS}\approx0.141~\mathrm{ms}$} (\eqrefalt{eq:t_dyn}).}
  \label{fig:bns_evolution_premerger}
\end{figure*}

\begin{figure*}[!h]
  \centering
  \begin{tabular}{c|c}
    Neutrino leakage \textbf{enabled} & Neutrino leakage \textbf{disabled}\\
    \subfloat[$t=102t_{\rm dyn,NS}=14.4~\mathrm{ms}$.]{
      \includegraphics[width=0.45\textwidth,clip]{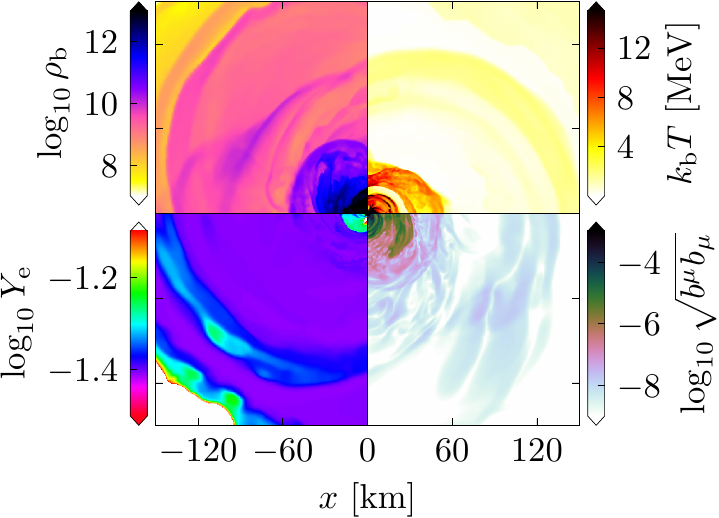}
      \label{fig:bns_leak4}
    }
    &
    \subfloat[$t=102t_{\rm dyn,NS}=14.4~\mathrm{ms}$.]{
      \addtocounter{subfigure}{2}
      \includegraphics[width=0.45\textwidth,clip]{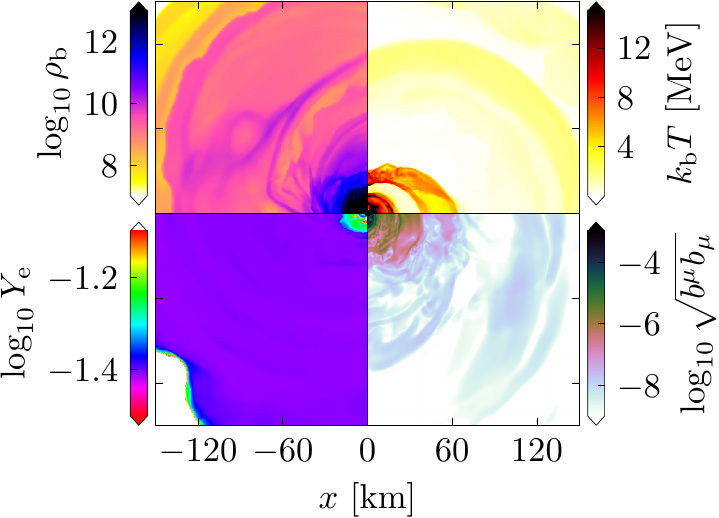}
      \label{fig:bns_noleak4}
    }
    \\
    \subfloat[$t=136t_{\rm dyn,NS}=19.2~\mathrm{ms}$.]{
      \addtocounter{subfigure}{-3}
      \includegraphics[width=0.45\textwidth,clip]{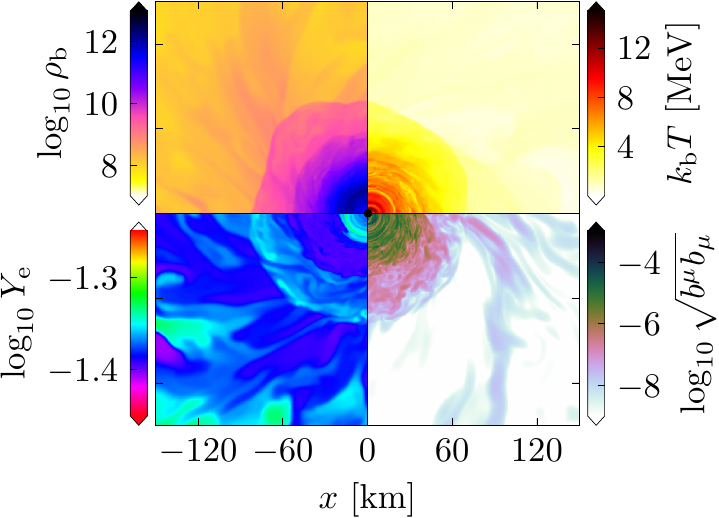}
      \label{fig:bns_leak5}
    }
    &
    \subfloat[$t=136t_{\rm dyn,NS}=19.2~\mathrm{ms}$.]{
      \addtocounter{subfigure}{2}
      \includegraphics[width=0.45\textwidth,clip]{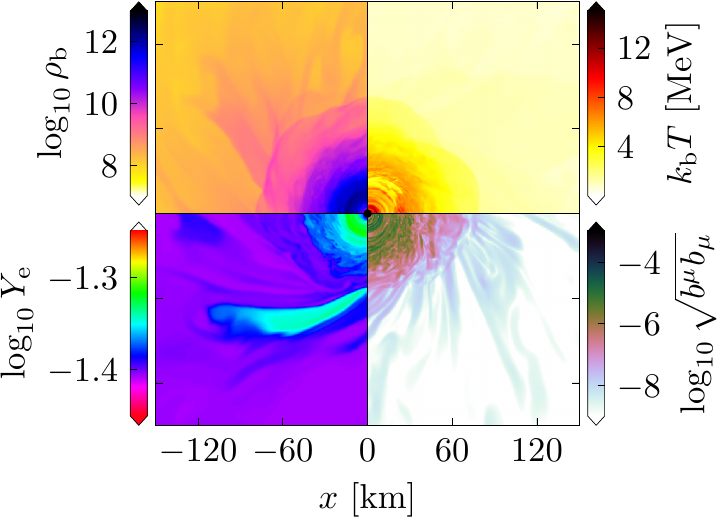}
      \label{fig:bns_noleak5}
    }
    \\
    \subfloat[$t=170t_{\rm dyn,NS}=24~\mathrm{ms}$.]{
      \addtocounter{subfigure}{-3}
      \includegraphics[width=0.45\textwidth,clip]{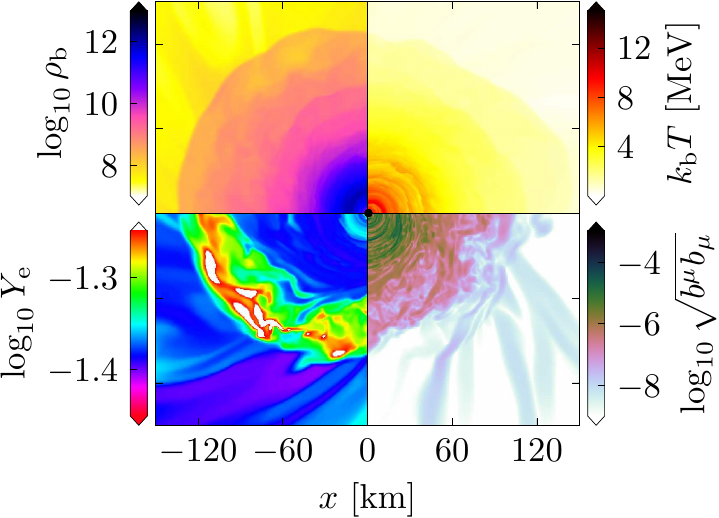}
      \label{fig:bns_leak6}
    }
    &
    \subfloat[$t=170t_{\rm dyn,NS}=24~\mathrm{ms}$.]{
      \addtocounter{subfigure}{2}
      \includegraphics[width=0.45\textwidth,clip]{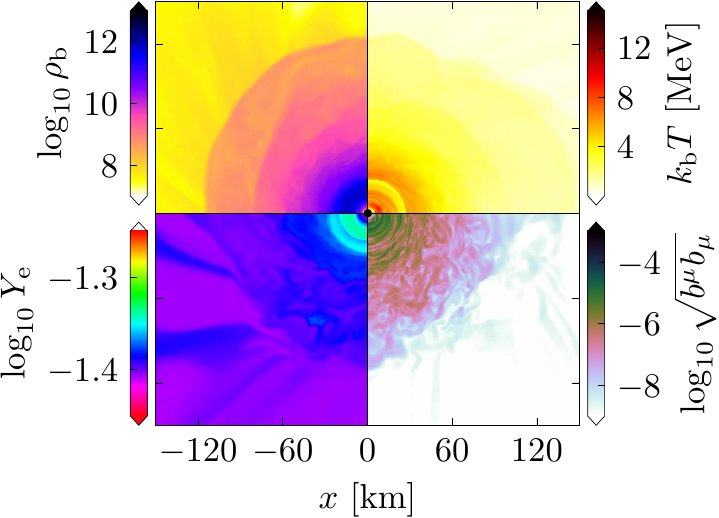}
      \label{fig:bns_noleak6}
    }
  \end{tabular}
  \caption{Same as figure \figref{fig:bns_evolution_premerger} for the post-merger phase.
  }
  \label{fig:bns_evolution_postmerger}
\end{figure*}

The gravitational wave signal already provides a hint that the inspiral phase of both simulations are virtually the same, and we confirm this behavior by looking at other quantities. In \figref{fig:bns_evolution_premerger} snapshots of density, temperature, electron fraction, and $b^{2}=b^{\mu}b_{\mu}$ are plotted on the orbital plane during the inspiral for the simulations with and without neutrino leakage. Indeed we observe no significant differences throughout.

However, the post-merger phase of these simulations exhibit noticeable differences, as shown in \figref{fig:bns_evolution_postmerger}. The BH accretion disk is slightly less neutron rich in the simulation with neutrino leakage enabled, with a large region of larger electron fraction being visible in the outer layers of the accretion disk. Notably, the same behavior is observed in \harmnuc when studying magnetized BH accretion disks (see Fig.~11 of~\cite{Murguia-Berthier:2021tnt}). \add{A detailed analysis of the global quantities that characteriza the disk evolution (e.g., mass, radius, accretion rate, average \ye, etc) will be the subject of a future work.}

In this study we are only presenting results up to ${\approx}5~\mathrm{ms}$ after black hole formation, when the spacetime already appears to be sufficiently static. We plan to use the recently developed \handoff code to transfer simulation data from \igm to \harmnuc and continue the post-merger phase for $\mathcal{O}\left({\sim}\mathrm{seconds}\right)$. Results of these continuation runs will be reported in a future publication.

\section{Conclusions \& Future Work}
\label{sec:conclusions}

Modeling magnetized compact binary systems in particular, and magnetized fluid flows in general, is of paramount importance for multimessenger astronomy. Simulating these systems accurately and reliably may not only lead to insights on phenomena we do not yet fully understand, but also provide crucial reference points for detections of gravitational waves and their electromagnetic and/or neutrino counterparts.

Given the importance of these simulations, many groups have developed GRMHD codes capable of performing them. Among these codes, the original GRMHD code developed by the Illinois numerical relativity group is particularly notable for its reliability and robustness when simulating a very broad range of astrophysical phenomena. Since \igm acts as an open-source, drop-in replacement of the original code, it inherits all of the original code's qualities while being faster and more concise.

The new version of \igm presented in this work aims at improving not only technical aspects of the code, but also the physical realism of the simulations that it can perform. To this end, two key new features were added: support for microphysical, finite-temperature, tabulated EOSs via a new \nrpy-based code---\nrpyeos; and neutrino physics via a leakage scheme using another \nrpy-based code---\nrpyleakage.

\igm has been developed to facilitate widespread community adoption. To this end, it was designed to be user-friendly, modular/extensible, robust, and performant/scalable. The development of this new version of the code shares all of these core principles, and it is with them in mind that we are making the code open-source and freely available for download~\cite{igm_github}.

In terms of user-friendliness, the code is well-documented, properly commented, and requires only basic programming skills to understand and run, traits also shared by \nrpyeos and \nrpyleakage. In the near future, we will release a series of \jupyter notebooks that meticulously document all of these codes.

Designed as thorns for the \etk and with clear separation of key algorithms in mind, all of these codes are both modular and extensible. To preserve the robustness of \igm, every new addition has been rigorously tested to ensure maximum reliability and optimal performance. A systematic study of the scalability of the new version of \igm, however, is not a focus of this work, in part because the core AMR infrastructure in the \etk is undergoing a major upgrade that will greatly improve scalability. \igm will be made compatible with this updated infrastructure in the coming months.

It is widely known that moving from a simple, analytic EOS to a tabulated EOS and adding a neutrino leakage scheme negatively affects the code's overall performance. In the case of \igm, the new version of the code is about $1.8\times$ slower than the hybrid equation of state version. This performance impact is comparable to those observed by the authors of other codes, such as \spritz. A code comparison study that showcases the performance and impact of different algorithmic choices made by different GRMHD codes will be the subject of future work.

The BNS results presented in this paper exist as a part of a larger project. \add{Because of this, we have taken special care to ensure that all of the new features in \igm are consistent with those in \harmnuc~\cite{Murguia-Berthier:2021tnt}.} The simulation data will be transferred from \igm to \harmnuc (a code specially designed to accurately and reliably model BH accretion disks) using our recently developed \handoff code~\cite{Armengol:2021mbt}, \rem{in which data from \igm is transferred to \harmnuc,} and the simulation \add{will} continue\rem{d} for $\mathcal{O}\left({\sim}\mathrm{seconds}\right)$. Results of these simulations will be presented in a future paper.

\section*{Acknowledgments}
The authors would like to thank E.~O'Connor for his comments on how nucleon-nucleon Bremsstrahlung is implemented in \groned. We would also like to thank M.~Campanelli, Y.~Zlochower, and T.~Piran for useful discussions\xspace\rep{and suggestions}{, as well as the anonymous referees for useful suggestions}. The plots in this paper have been generated using \mpl~\cite{Hunter:2007}; all plotting scripts can be made available upon request. This work was primarily funded through NASA award TCAN-80NSSC18K1488. L.R.W.\ and Z.B.E.\ gratefully acknowledge support from NSF awards PHY-1806596, PHY-2110352, OAC-2004311, as well as NASA award ISFM-80NSSC18K0538. \add{The material is based upon work supported by NASA under award number 80GSFC21M0002.} A.M-B.\ is supported by NASA through the NASA Hubble Fellowship grant HST-HF2-51487.001-A awarded by the Space Telescope Science Institute, which is operated by the Association of Universities for Research in Astronomy, Inc., for NASA, under contract NAS5-26555. This research made use of Idaho National Laboratory computing resources which are supported by the Office of Nuclear Energy of the U.S. Department of Energy and the Nuclear Science User Facilities under Contract No. DE-AC07-05ID14517, as well as TACC’s Frontera NSF projects PHY20010 and AST20021. Additional resources were provided by the RIT's Green Pairies Cluster, acquired with NSF MRI grant PHY1726215.

\bibliographystyle{apsrev4-1}
\bibliography{references}

\end{document}